\documentclass{article}

\usepackage{amssymb}
\usepackage{amsmath}
\usepackage{epsfig}
\renewcommand{\theequation}{\arabic{section}.\arabic{equation}}

\begin{document}

\title{\bf Effects of Schwarzschild Black Hole Horizon on Isothermal Plasma Wave Dispersion}

\author{M. Sharif \thanks{msharif@math.pu.edu.pk} and Umber Sheikh\\
%EndAName
\textit{\ {\small Department of Mathematics,}}\\
\textit{{\small University of the Punjab, Lahore 54590,
Pakistan}}}

 \date{}
  \maketitle
\begin{abstract}
The 3+1 GRMHD equations for Schwarzschild spacetime in Rindler
coordinates with isothermal state of plasma are formulated. We
consider the cases of non-rotating and rotating backgrounds with
non-magnetized and magnetized plasmas. For these cases, the
perturbed form of these equations are linearized and Fourier
analyzed by introducing plane wave type solutions. The determinant
of these equations in each case leads to two dispersion relations
which give value of the wave number $k$. Using the wave number, we
obtain information like phase and group velocities etc. which help
to discuss the nature of the waves and their characteristics. These
provide interesting information about the black hole magnetosphere
near the horizon. There are cases of normal and anomalous
dispersion. We find a case of normal dispersion of waves when the
plasma admits the properties of Veselago medium. Our results agree
with those of Mackay et al. according to which rotation of a black
hole is required for negative phase velocity propagation.
\end{abstract}


\section{Introduction}

General Relativity (GR) is a beautiful scheme for describing the
gravitational field and the corresponding equations. This theory is
believed to apply to all forms of interaction whether these are
between the charged particles or large scale gravitational
structures, including compact objects like black holes. Black holes
are still mysterious \cite{VS}. While physicists have been grappling
the theory of black holes, astronomers have been searching for
real-life examples of black holes in the universe \cite{Na}. It is
proved that black holes exist on the basis of study of effects which
they exert on their surroundings. With their enormous gravitational
fields, they greatly affect the surrounding plasma medium (which is
highly magnetized) and hence plasma physics in the vicinity of a
black hole has become a subject of interest in astrophysics. In the
immediate vicinity of a black hole, GR applies and it is therefore
of interest to formulate plasma physics problem in this context.

In large scale gravitational structures, one often encounters
magnetic fields. All the massive stars, neutron stars and black
holes carry energy flux which produces a relatively large magnetic
field. In order to understand such phenomena, we have a developed
theory of magnetized plasma called theory of magnetohydrodynamics
(MHD). In this theory, perfect MHD flow is important which is the
flow of plasma with negligible charge separation and with a magnetic
field frozen into it \cite{Be}. Since all the compact objects have
strong gravitational fields near their surfaces \cite{Pe}, it is
important to study the general relativistic effects on physical
processes taking place in their vicinity. The MHD theory with the
effects of gravity is called general relativistic
magnetohydrodynamics (GRMHD). The GRMHD equations include Maxwell's
equations, Ohm's law, mass, momentum and energy conservation
equations. These help us to study stationary configurations and
dynamic evolution of conducting fluid in a magnetosphere. They are
also required to investigate various aspects of the interaction of
relativistic gravity with plasma's magnetic field.

The general relativistic physics of black holes and plasmas takes
place in 3+1 dimensional space while the corresponding relativistic
laws are written in three-dimensional language. This formalism is
well-suited to carry non-relativistic intuition of physicists about
plasmas, hydrodynamics and stellar dynamics into the arena of black
holes and GR. The formalism (also called ADM formalism) was
originally developed by Arnowitt et al. \cite{ADM} and was motivated
by several startling results proved in 1970 using the black hole
viewpoint \cite{Ha}-\cite{Da}. Thorne and Macdonald
\cite{TM1}-\cite{TPM} extended the formulation to electromagnetic
fields of the black hole theory. The wave propagation theory in the
Friedmann universe was investigated using 3+1 formalism by Holcomb
and Tajima \cite{HT}, Holcomb \cite{Ho} and Dettmann et al.
\cite{De}. Khanna \cite{Kh} derived the MHD equations describing the
two component plasma theory of the Kerr black hole in this split.
Ant\'{o}n et al. \cite{A} used the formalism to investigate various
test simulations and discussed magneto-rotational instability of
accretion disks. Komissarov \cite{Ko1} discussed Blendford-Znajec
monopole solution using 3+1 formalism in black hole electrodynamics.
Zhang \cite{Z1}-\cite{Z2} formulated the black hole theory for
stationary symmetric GRMHD using this formalism with its
applications in Kerr geometry.

The gravitational perturbations away from the Schwarzschild
background have been discussed by Regge and Wheeler \cite{RW}. This
treatment was extended by Zerilli \cite{Ze} who showed that the
perturbations corresponding to a change in mass, the angular
momentum and charge of the Schwarzschild black hole are
well-behaved. The decay of non well-behaved perturbations has been
investigated by Price \cite{Pr}. The quasi-static electric problem
was solved by Hanni and Ruffini \cite{HR} who proved that the lines
of force diverge at the horizon for the observer at infinity. Wald
\cite{Wa} derived the solution for electromagnetic field occurring
when a stationary axisymmetric black hole is placed in a uniform
magnetic field aligned along the symmetry axis of black hole. A
linearized treatment of plasma waves for special relativistic
formulation of the Schwarzschild black hole was developed by Sakai
and Kawata \cite{SK}. Using 3+1 ADM formalism, Buzzi et al.
\cite{Bu} extended this treatment to waves in general relativistic
two component plasma propagating in radial direction. They
investigated the one dimensional radial propagation of transverse
and longitudinal waves close to the Schwarzschild horizon. In a
recent paper, Sharif and Umber \cite{MS} have found some interesting
properties of cold plasma in the vicinity of a Schwarzschild black
hole horizon using 3+1 GRMHD equations. They have also verified that
the rotation is a cause of negative phase velocity propagation which
coincides with the result of Mackay et al. \cite{Mackay}.

The aim of this paper is to discuss the properties of waves in
isothermal plasma. We develop and expound a theoretical method
relevant to the study of relativistic fluid model used in
applications to astrophysics and plasma physics. We consider
Schwarzschild spacetime in Rindler coordinates which gives an
example of the essential ideas of the horizon and the 3+1 split
without distracting complications of curved spacial three geometry.
This investigation would lead to support the viewpoint that
information from a black hole cannot be extracted.

The GRMHD equations in 3+1 formalism using isothermal plasma in the
Schwarzschild magnetosphere are obtained. The component form of the
equations is extracted by involving linear perturbations in each of
the cases of either non-magnetized or magnetized, non-rotating or
rotating perfect fluid. The condition for the plane harmonic wave is
imposed on all the perturbations. The equations are then Fourier
analyzed and solved to get the dispersion relations with the help of
software \emph{Mathematica}. Two dispersion relations are found in
each case and the wave numbers are deduced uniquely in terms of
angular frequency. The wave velocity, group velocity, refractive
index and the corresponding quantities give the properties of plasma
near the horizon expressed by the perfect fluid. These properties
are shown in terms of graphs.

The scheme of the paper is as follows. The next section contains
basic informatory results for 3+1 formalism. The results for the
GRMHD equation are given and energy equation is obtained. These
equations are restricted for the case of perfect fluid with perfect
MHD flow assumption for Rindler approximation of the Schwarzschild
spacetime. The equations are further modified by adding the
condition of isothermal plasma. Section $3$ is devoted to the case
of non-rotating background for which we find the dispersion
relations. Section $4$ is furnished with the values of $k$ for
rotating non-magnetized plasma. In section $5$, we generalize the
problem to the case of rotating magnetized plasma. In the last
section, we shall summarize and discuss the results.

\section{3+1 Spacetime Modeling}

The astrophysical theory in early 1960s was dominated by the frozen
star viewpoint \cite{Zel}.  As long as the viewpoint prevailed,
physicists failed to realize that the black hole can be dynamical,
evolving, energy-releasing object. Many problems of current interest
in black-hole astrophysics cannot be posed in frozen-star viewpoint,
because the viewpoint has difficulty producing unambiguous boundary
conditions at the event horizon. Such problems where it is
inadequate to replace the black hole horizon, in picture and
calculations, by a surface of no return, a new viewpoint for the
black hole physics was developed - \emph{the membrane paradigm} (a
marriage of horizon formalism with 3+1 formalism). This is
mathematically equivalent to the standard, full, general
relativistic theory of black holes for the physics outside the
horizon. Although this formalism is not treated in standard
relativity textbooks, it has been developed in great detail by many
researchers and has played an important role \cite{TPM}.

In $3+1$ formalism, the line element of the spacetime can be
written as \cite{Z2}
\begin{equation*}
ds^2=-\alpha^2dt^2+\gamma_{ij}(dx^i+\beta^idt)(dx^j+\beta^jdt),
\end{equation*}
where $\alpha$ is the lapse function, $\beta^i$ are the components
of shift vector and $\gamma_{ij}$ are the components of the
spatial metric. All these quantities are functions of the
coordinates $t,~x^i$. A natural observer, associated with this
spacetime called fiducial observer (FIDO), has four-velocity
\textbf{n} perpendicular to the hypersurfaces of constant time
$t$. Notice that we are using geometrized units.

The Rindler approximation of the Schwarzschild line element is
\cite{TPM}
\begin{equation}
ds^2=-\alpha^2(z)dt^2+dx^2+dy^2+dz^2
\end{equation}
which is an analogue of the Schwarzschild spacetime with $z$, $x$
and $y$ as radial $r$, axial $\phi$ and poloidal $\theta$ directions
respectively. In Rindler coordinates, the lapse function $\alpha$ is
$\frac{z}{2r_H}$, where $r_H$ is the radius of the Schwarzschild
black hole. This function clearly vanishes at the horizon which we
can place at $z=0$. It is useful to think of this absolute space
with fiducial observer which never moves from its fixed location.
The relationship between the universal time $t$ and FIDO measured
time $\tau$ can be expressed by the time lapse
$\alpha\equiv\frac{d\tau}{dt}$. The Schwarzschild black hole is
non-rotating, hence the shift vector $\beta=0$. This is the minimum
configuration of the Kerr black hole out of which no energy can be
extracted.

\subsection{Perfect GRMHD Equations}

In magnetohydrodynamic treatment, the plasma is represented as a
perfect fluid. The perfect MHD flow assumption is given by
\begin{equation}\label{pM}
\textbf{E}+\textbf{V}\times\textbf{B}=0,
\end{equation}
where $\textbf{V},~\textbf{B}$ and $\textbf{E}$ are the velocity,
magnetic and electric fields of the fluid measured by FIDO. This
assumption shows that fluid possesses no electric field in
rest-frame and the electric field is perpendicular to the magnetic
field.

It is known that unlike cold plasma approximation, isothermal plasma
contains plasma pressure which cannot be ignored. The pressure
gradient term is to be included. In this model, the energy equation
is also introduced but the term involving the heat flux is
neglected. Because of the vanishing thermal conductivity, the plasma
is non-viscous and consequently the non-diagonal terms of the
pressure dyad are all equal to zero. FIDO measured local energy
conservation law \cite{Z1} is given by
\begin{equation}\label{1}
\frac{d\epsilon}{d\tau}+\theta
\epsilon+\frac{1}{2\alpha}W^{ij}(\pounds_t \gamma^{ij})
=-\frac{1}{\alpha^2}\nabla.(\alpha^2
\bf{S})+\frac{1}{\alpha}\nabla\beta:
\overleftrightarrow{\bf{W}}+\bf{E}.\bf{j}.
\end{equation}
Here $\epsilon,~\textbf{S},~W^{ij}$ and $\bf{j}$ represent the mass
energy density, energy flux, stress tensor and current vector of the
electromagnetic field in three-dimensions respectively, $\theta$ is
the expansion rate of the FIDO's four-velocity, $\pounds_t$ is the
time derivative along shifting congruence (Lie derivative with
respect to global time in a usual fashion) and $\frac{d}{d
\tau}\equiv \frac{1}{\alpha}(\frac{\partial}{\partial t
}-\beta.\nabla)$ is the rate of change of a three-dimensional vector
which lies in the absolute space according to the FIDO. When we
substitute the following values for perfect fluid \cite{TM1}
\begin{eqnarray*}
\epsilon&=&(\mu \rho_0-p(1-\textbf{V}^2))\gamma^2,\\
\textbf{S}&=&\mu\rho_0\gamma^2\textbf{V},\\
\overleftrightarrow{\textbf{W}}&=&\mu\rho_0\gamma^2\textbf{V}
\otimes\textbf{V}+p\overleftrightarrow{\gamma}
\end{eqnarray*}
with $\mu$ the specific enthalpy, $\rho_0$ is the rest-mass density,
$p$ is the pressure, and perfect MHD condition, Eq.(\ref{1}) takes
the following form
\begin{eqnarray}\label{2}
&&\rho_0\gamma^2\frac{D\mu}{D\tau}+\mu\gamma^2\frac{D\rho_0}{D\tau}
+2\rho_0\mu\gamma^4\textbf{V}.\frac{D\textbf{V}}{D\tau}-\frac{d
p}{d\tau}+(\mu\rho_0\gamma^2-p)\theta+2\rho_0\mu\gamma^2\textbf{V}.\textbf{a}
+\rho_0\mu\gamma^2(\nabla.\textbf{V})\nonumber\\
&&-\frac{1}{\alpha}\{\rho_0\mu\gamma^2\textbf{V}.(\textbf{V}.\nabla)\beta
+p(\nabla.\beta)\}+\frac{1}{2\alpha}\{\rho_0
\mu\gamma^2V^iV^j+p\gamma^{ij}\}\pounds_t\gamma_{ij}
+\frac{1}{4\pi\alpha}[(\textbf{V}\times\textbf{B}).(\nabla
\times (\alpha\textbf{B}))\nonumber\\
&&+\alpha(\textbf{V}\times\textbf{B}).\frac{d}{d\tau}(\textbf{V}\times\textbf{B})
+(\textbf{V}\times\textbf{B}).(\textbf{V} \times
\textbf{B}.\nabla)\beta+\theta\alpha(\textbf{V}\times\textbf{B}).
(\textbf{V}\times\textbf{B})]=0,
\end{eqnarray}
where $$\frac{D}{D\tau}\equiv\frac{d}{d\tau}+\textbf{V}.\nabla
=\frac{1}{\alpha}\left\{\frac{\partial}{\partial
t}+(\alpha\textbf{V}-\beta).\nabla\right\}.$$

For the Schwarzschild black hole, FIDO measured Faraday's law,
equation of evolution of the magnetic field, local conservation laws
of mass and momentum are given by Eqs.(2.17)-(2.20) \cite{MS}. These
equations along with local conservation law of energy Eq.(\ref{2})
become
\begin{eqnarray}
&&\frac{\partial \textbf{B}}{\partial t}=\nabla
\times(\alpha \textbf{V}\times \textbf{B}),\\
&&\frac{\partial \textbf{B}}{\partial t}+(\alpha
\textbf{V}.\nabla)\textbf{B}-(\textbf{B}.\nabla)(\alpha \textbf{V}
)+(\nabla.(\alpha \textbf{V}))\textbf{B}=0,\\
&&\frac{\partial (\rho_0 \mu)}{\partial t}+\{(\alpha
\textbf{V}).\nabla\}(\rho_0 \mu)+\rho_0 \mu
\gamma^2\textbf{V}.\frac{\partial \textbf{V}}{\partial t}+\rho_0
\mu \gamma^2\textbf{V}.(\alpha \textbf{V}.\nabla)\textbf{V}+
\rho_0 \mu \{\nabla.(\alpha\textbf{V})\}=0,\\
&&\left\{\left(\rho_0\mu
\gamma^2+\frac{\textbf{B}^2}{4\pi}\right)\delta_{ij}+\rho_0\mu
\gamma^4
V_iV_j-\frac{1}{4\pi}B_iB_j\right\}\left(\frac{1}{\alpha}\frac{\partial}{\partial
t}+\textbf{V}.\nabla\right)V^j-\left(\frac{\textbf{B}^2}{4\pi}\delta_{ij}
-\frac{1}{4\pi}B_iB_j\right){V^j}_{,k}V^k\nonumber\\
&&+\rho_0 \gamma^2 V_i\left\{\frac{1}{\alpha}\frac{\partial
\mu}{\partial t}+(\textbf{V}.\nabla)\mu\right\}=-\rho_0\mu
\gamma^2 a_i-p_{,i}+ \frac{1}{4\pi}(\textbf{V}\times \textbf{B})_i
\nabla.(\textbf{V}\times \textbf{B})
-\frac{1}{8\pi\alpha^2}(\alpha \textbf{B})^2_{,i}\nonumber\\
&&+\frac{1}{4\pi\alpha}(\alpha
B_i)_{,j}B^j-\frac{1}{4\pi\alpha}[\textbf{B}\times\{\textbf{V}
\times (\nabla \times (\alpha \textbf{V} \times
\textbf{B}))\}]_i,\\
&&\gamma^2\left(\frac{1}{\alpha}\frac{\partial}{\partial
t}+\textbf{V}.\nabla\right)(\mu\rho_0)-\frac{1}{\alpha}\frac{\partial
p}{\partial
t}+2\rho_0\mu\gamma^4\textbf{V}.\left(\frac{1}{\alpha}\frac{\partial
}{\partial t}+\textbf{V}.\nabla\right)\textbf{V}
+2\rho_0\mu\gamma^2(\textbf{V}.\textbf{a})\nonumber\\
&&+\rho_0\mu\gamma^2(\nabla.\textbf{V})
+\frac{1}{4\pi\alpha}\left[(\textbf{V}\times\textbf{B}).(\nabla
\times(\alpha\textbf{B}))+(\textbf{V}\times\textbf{B}).\frac{\partial}{\partial
t}(\textbf{V}\times\textbf{B})\right]=0,
\end{eqnarray}
where $\textbf{a}$ is the gravitational acceleration. Notice that
here Eq.(\ref{2}) is the new equation in addition to
Eqs.(2.17)-(2.20) \cite{MS}. These constitute the perfect GRMHD
equations for fluid with non-zero pressure in the vicinity of a
black hole horizon.

\subsection{GRMHD Equations for Isothermal Plasma in Schwarzschild Spacetime}

The isothermal equation of state can be expressed by the following
equation \cite{Z1}
\begin{equation}\label{EOS}
\mu=\frac{\rho+p}{\rho_0}=constant.
\end{equation}
Here $\rho$ is the mass density and pressure of the fluid.

For the Rindler approximation of Schwarzschild geometry, FIDOs
four-velocity expansion rate and shift vector vanish. Thus the
perfect GRMHD equations in 3+1 formalism for isothermal state of
plasma turn out to be
\begin{eqnarray}\label{3}
&&\frac{\partial \textbf{B}}{\partial t}=\nabla \times(\alpha
\textbf{V} \times \textbf{B}),\\
\label{4}
&&\frac{\partial \textbf{B}}{\partial t}+(\alpha
\textbf{V}.\nabla)\textbf{B}-(\textbf{B}.\nabla)(\alpha
\textbf{V})+\textbf{B}\nabla.(\alpha \textbf{V})=0,\\
\label{5}
&&\frac{\partial (\rho+p)}{\partial t}+(\alpha
\textbf{V}.\nabla)(\rho+p)
+(\rho+p)\gamma^2\textbf{V}.\frac{\partial \textbf{V}}{\partial
t}+(\rho+p)\gamma^2\textbf{V}.(\alpha
\textbf{V}.\nabla)\textbf{V}+(\rho+p) \nabla.(\alpha\textbf{V})=0,\\
\label{6}
&&\left\{\left((\rho+p)\gamma^2+\frac{\textbf{B}^2}{4\pi}\right)\delta_{ij}
+(\rho+p)\gamma^4V_iV_j-\frac{1}{4\pi}B_iB_j\right\}
\left(\frac{1}{\alpha}\frac{\partial}{\partial
t}+\textbf{V}.\nabla\right)V^j\nonumber\\
&&-\left(\frac{\textbf{B}^2}{4\pi}\delta_{ij}-\frac{1}{4 \pi}B_i
B_j\right)V^j_{,k}V^k=-(\rho+p)\gamma^2 a_i-p_{,i}+\frac{1}{4\pi}
\nabla.(\textbf{V}\times\textbf{B})(\textbf{V}\times \textbf{B})_i\nonumber\\
&&-\frac{(\alpha \textbf{B})^2_{,i}}{8\pi\alpha^2}+\frac{(\alpha
B_i)_{,j}B^j}{4 \pi \alpha}-\frac{1}{4 \pi
\alpha}[\textbf{B}\times\{\textbf{V} \times
(\nabla \times (\alpha \textbf{V} \times \textbf{B}))\}]_i,\\
\label{7}
&&\gamma^2(\frac{1}{\alpha}\frac{\partial}{\partial
t}+\textbf{V}.\nabla)(\rho+p)
+2(\rho+p)\gamma^4\textbf{V}.(\frac{1}{\alpha}\frac{\partial}{\partial
t}+\textbf{V}.\nabla)\textbf{V}+2(\rho+p)\gamma^2\textbf{V}.\textbf{a}
-\frac{1}{\alpha}\frac{\partial p}{\partial t}\nonumber\\
&&+(\rho+p)\gamma^2\nabla.\textbf{V}+\frac{1}{4\pi\alpha}[(\textbf{V}
\times \textbf{B}).(\nabla \times (\alpha
\textbf{B}))+(\textbf{V}\times\textbf{B}).
\frac{\partial}{\partial t}(\textbf{V} \times \textbf{B})]=0.
\end{eqnarray}
Eqs.(\ref{3})-(\ref{7}) are the perfect GRMHD equations for
isothermal plasma in the vicinity of the Schwarzschild black hole.
In the rest of the paper, the Fourier analysis method will be
applied to the perturbed form of these equations (including the
restrictions of backgrounds) to obtain the dispersion relations.

\section{Non-Rotating Background}

In non-rotating background, we apply the Fourier analysis method to
the perturbed GRMHD equations given in Appendix A. These equations
will lead to the dispersion relations. We obtain solution of
dispersion relations as graphs. The same procedure will be applied
to rotating background.

We assume the plane wave type solutions of the form $\it{e^{-i
(\omega t-k z)}}$ for perturbations (Appendix A). Thus the perturbed
variables can be expressed as follows
\begin{eqnarray}{\setcounter{equation}{1}}
\label{3.1} \tilde{\rho}(t,z)=c_1e^{-\iota(\omega t-kz)},\quad
\tilde{p}(t,z)=c_2e^{-\iota(\omega t-kz)},\quad
v_z(t,z)=c_3e^{-\iota (\omega t-kz)},\quad b_z(t,z)=c_4e^{-\iota
(\omega t-kz)},
\end{eqnarray}
where $k$ is the wave number and $\omega$ is the angular frequency.

Using Eq.(\ref{3.1}) the Fourier analyzed form of the perturbed
Eqs.(\ref{3.7})-(\ref{3.11}) specified for the non-rotating
background, it follows that
\begin{eqnarray}\label{3.12}
&&-\frac{\iota \omega}{\alpha} c_4=0,\\
\label{3.13}
&&\iota k c_4=0,\\
\label{3.14}
&&c_1\{-\rho\iota \omega+\iota k\rho\alpha u-(u\alpha
p)'-\alpha u^2\gamma^2pu'\}+c_2\{-p\iota \omega+\iota kp\alpha
u+(u\alpha
p)'+\alpha u^2\gamma^2pu'\}\nonumber\\
&&+c_3(\rho+p)\left\{\alpha(1+\gamma^2u^2)\iota
k-\alpha(1-2\gamma^2u^2)(1+\gamma^2u^2)\frac{u'}{u}-i\omega\gamma^2
u\right\}=0,\\
\label{3.15}
&&c_1\rho\gamma^2 \{a_z+uu'(1+\gamma^2
u^2)\}+c_2\{p\gamma^2
\{a_z+uu'(1+\gamma^2 u^2)\}+\iota k p+p'\}\nonumber\\
&&+c_3(\rho+p)\gamma^2\left[(1+\gamma^2u^2)\left(\frac{-\iota
\omega}{\alpha}+\iota u k\right)+\{u'(1+\gamma^2 u^2)(1+4\gamma^2
u^2) +2\gamma^2 ua_z\}\right]=0,\\
\label{3.16}
&&\left\{\rho\gamma^2\left(\frac{-\iota\omega}{\alpha}+\iota
ku\right)
+\gamma^2u\rho'+\rho2\gamma^2ua_z+\rho(1+2\gamma^2u^2)\gamma^2u'\right\}c_1\nonumber\\
&&+\left\{\frac{-\iota\omega}{\alpha}p(\gamma^2-1)+\iota
k\gamma^2up+\gamma^2up'+2\gamma^2upa_z+p(1+2\gamma^2u^2)\gamma^2u'\right\}c_2\nonumber\\
&&+(\rho+p)\gamma^2\left\{\frac{-2\iota\omega}{\alpha}\gamma^2u+(\iota
k+3\gamma^2uu'+a_z)(1+2\gamma^2u^2)-\frac{u'}{u}\right\}c_3=0.
\end{eqnarray}
Eqs.(\ref{3.12}) and (\ref{3.13}) yield that $c_4=0$, i.e., the
magnetic field has no effect of black hole gravity as well as time.
We would like to mention here that Eqs.(\ref{3.14})-(\ref{3.16})
also give non-magnetized plasma.

\subsection*{Numerical Solutions}

We consider that the magnetosphere is filled with stiff fluid for
which $\rho=constant=p$. We also assume that the time lapse
$\alpha=z$. Using these values, the mass conservation law in
3-dimensions, i.e., $\alpha(\rho+p)\gamma u=constant$  gives
$u=\frac{1}{\sqrt{1+z^2}}$. Due to the magnetic field parallel to
the wave number and the use of the longitudinal part of momentum
equation, the isothermal plasma shows only the longitudinal
(electron or ion) plasma waves.

When we use these values, the determinant of the coefficients of
constants $c_1,~c_2$ and $c_3$ in Eqs.(\ref{3.14})-(\ref{3.16})
leads to a complex dispersion equation. On comparing the real and
imaginary parts we obtain two dispersion equations. The real part
shows an equation of the type
$A_1(z)k^2+A_2(z,\omega)k+A_3(z,\omega)=0$ which on solving gives
two values of the wave number $k$. The imaginary part gives the
dispersion equation of the type
$A_1(z)k^3+A_2(z,\omega)k^2+A_3(z,\omega)k+A_4(z,\omega)=0$ which
provides three (including two complex conjugate) values of the wave
number. Using the values of $k$, the quantities like phase velocity
($v_p=\frac{\omega}{k}$), refractive index ($n=\frac{1}{v_p}$), its
change with respect to angular frequency ($\frac{dn}{d\omega}$) and
group velocity
($v_g=\frac{d\omega}{dk}=\frac{1}{n+\omega\frac{dn}{d\omega}}$) can
be obtained. These quantities help us to investigate the properties
of the waves as well as their dispersion.

The two dispersion relations obtained from the real part give the
same value of $k$. The corresponding graphs are given in Figure 1.
The wave number obtained from the imaginary part is shown in Figure
2.
\begin{figure}
\center \epsfig{file=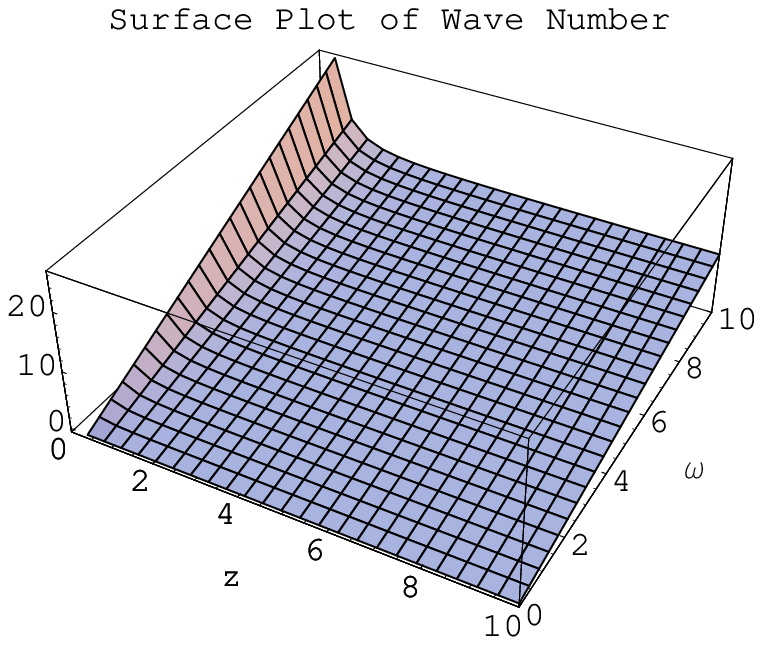,width=0.4\linewidth} \center
\epsfig{file=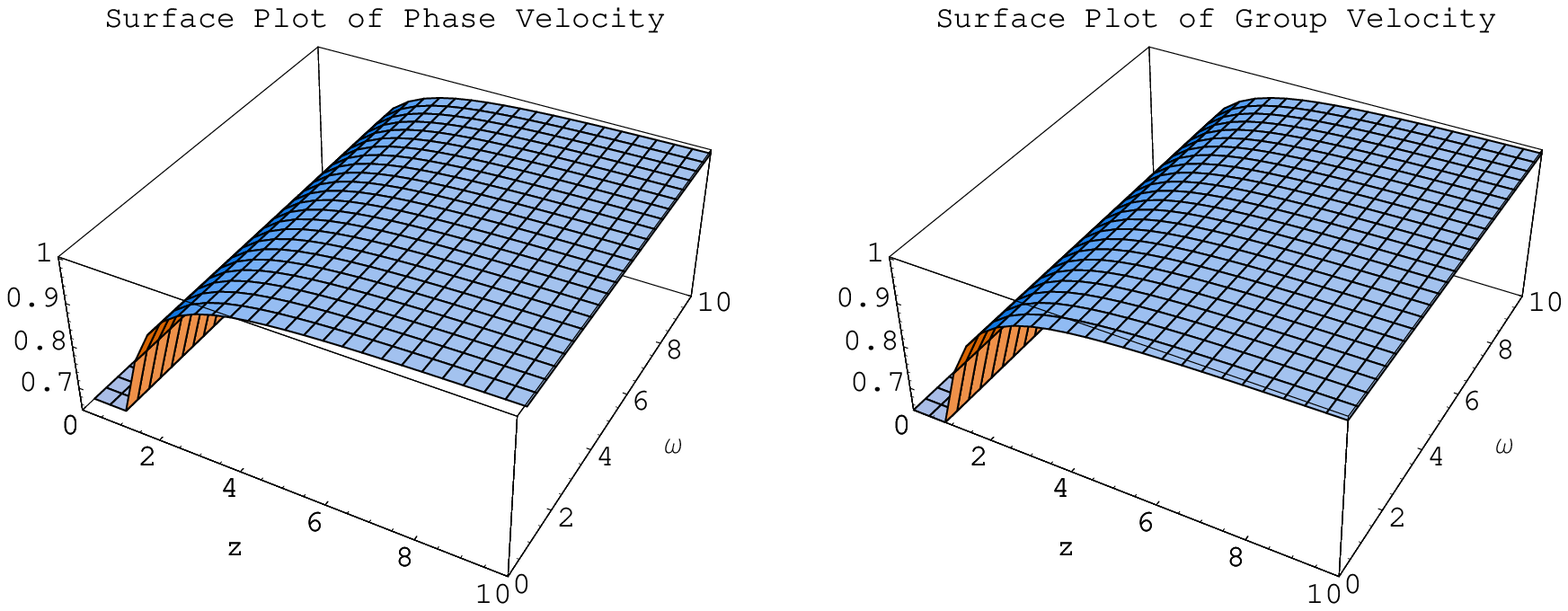,width=0.8\linewidth} \center
\epsfig{file=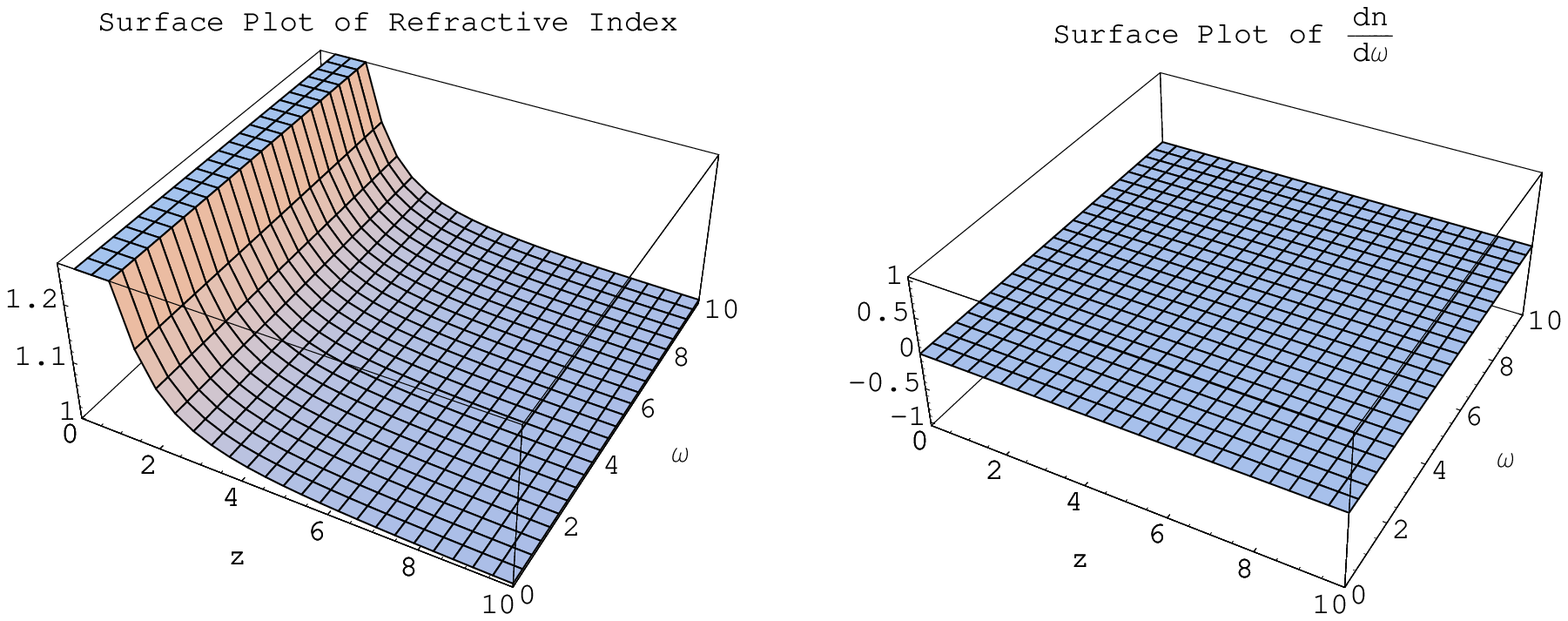,width=0.8\linewidth} \caption{The waves
decrease as they move away from the horizon. The phase and group
velocities are equal. The dispersion is not normal.}
\end{figure}

In Figure 1, the wave number $k$ becomes infinite at $z=0$ which
means that the waves vanish at the horizon due to effect of immense
gravity. No wave with zero angular frequency exist which is a
physical consequence. The wave number decreases as we go away from
the horizon which means that the waves lose energy as their distance
from the horizon increases. The wave number increases with the
increase in angular frequency. The phase and group velocities are
positive and increase when we go away from the horizon. We can
deduce that the increase in gravity decreases the wave velocity as
well as the speed with which energy travels. Both the phase and
group velocities are equal. The refractive index is greater than one
and there is no change with respect to angular frequency hence the
dispersion is not normal there \cite{Ja}.

\begin{figure}
\center \epsfig{file=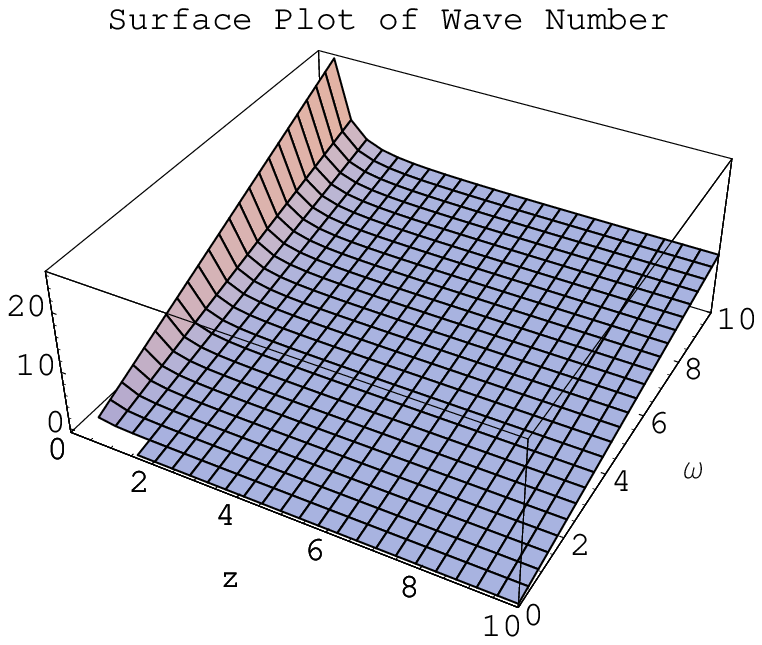,width=0.4\linewidth} \center
\epsfig{file=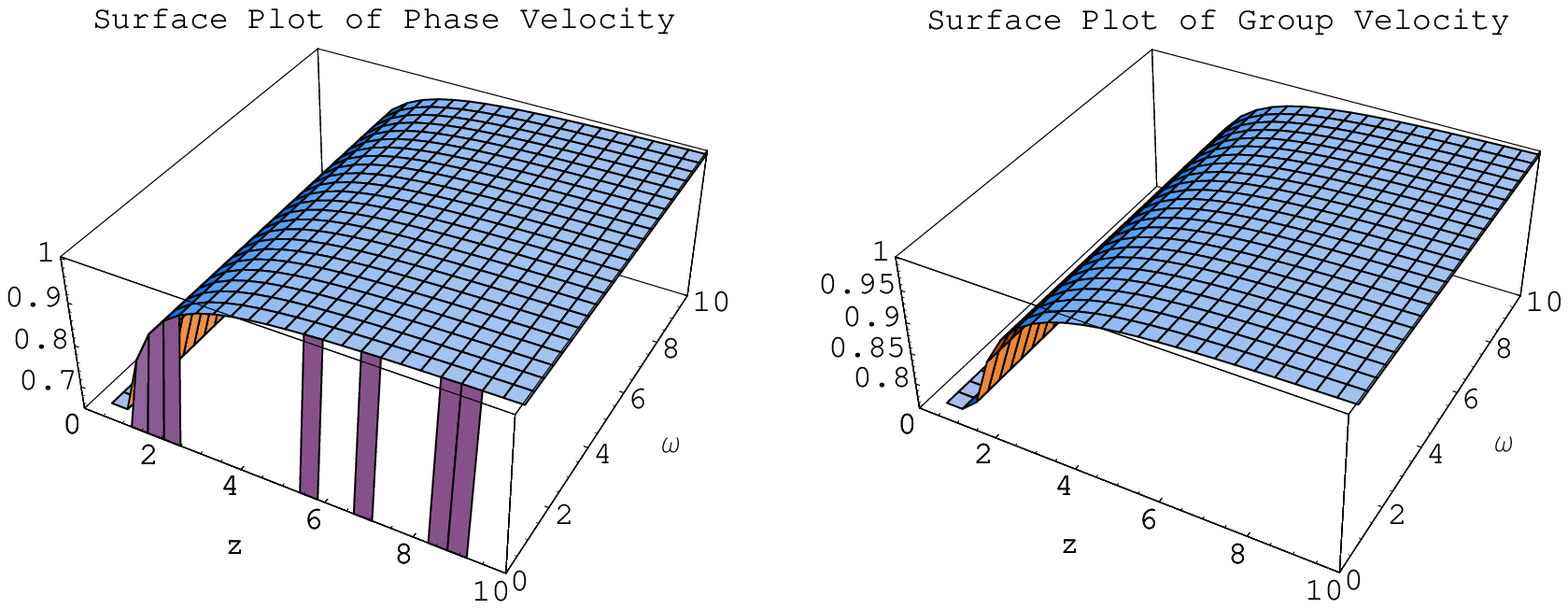,width=0.8\linewidth} \center
\epsfig{file=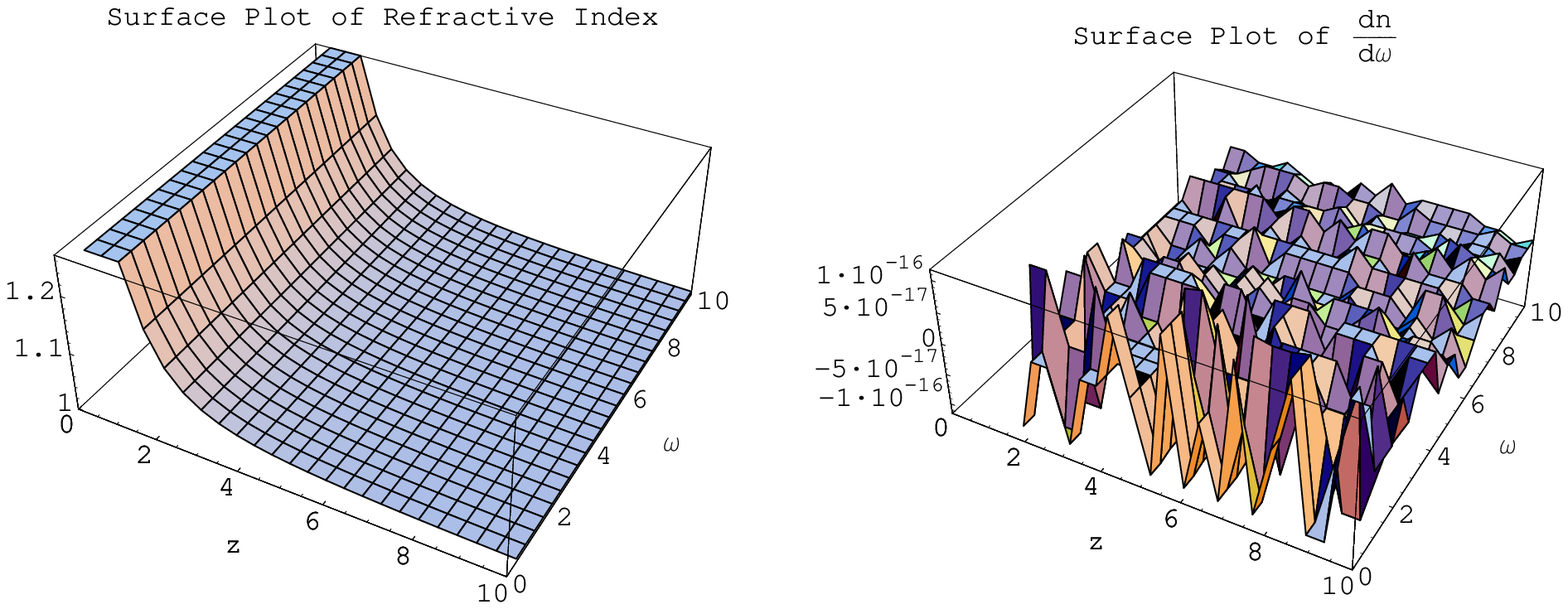,width=0.8\linewidth}\caption{The wave number
decreases as the waves move away from the event horizon. In most
of the region the phase and group velocities are equal and the
dispersion is not normal.}
\end{figure}
Figure 2 shows that the wave number is infinite at $z=0.$ In the
region $0<z<1,~0\leq\omega<6\times 10^{-6}$, the wave number takes
some complex value and the waves are evanescent there, otherwise
the wave number $k$ is positive. It decreases as we move away from
the event horizon of the Schwarzschild black hole. The increase in
angular frequency increases the wave number. The phase and group
velocities are equal except in the region $0<\omega<6\times
10^{-6}$. Since the phase velocity increases as we go away from
the horizon, the waves gain energy as they go far away from the
event horizon of the black hole. Refractive index is positive and
greater than one but its derivative with respect to angular
frequency has random variation. Hence the medium is not of normal
dispersion.

\section{Rotating Non-Magnetized Background}

The perturbed form of GRMHD Eqs.(\ref{3})-(\ref{7}) meant for
rotating non-magnetized background are given in Appendix B. For the
sake of Fourier analysis, we assume that the perturbations (Appendix
B) are of the plane wave form, i.e., $\it{e^{-i(\omega t-k z)}}$.
Thus the perturbed variables take the form
\begin{eqnarray}{\setcounter{equation}{1}}\label{4.4}
\tilde{\rho}(t,z)=c_1e^{-\iota (\omega t-kz)},\quad
\tilde{p}(t,z)=c_2e^{-\iota (\omega t-kz)},\quad
v_z(t,z)=c_3e^{-\iota (\omega t-kz)},\quad v_x(t,z)=c_4e^{-\iota
(\omega t-kz)}.
\end{eqnarray}

Using Eq.(\ref{4.4}), we can write the Fourier analyzed
Eqs.(\ref{4.8})-(\ref{4.11}) for rotating non-magnetized background
as follows
\begin{eqnarray}\label{4.12}
&&c_1\{(-\iota\omega+\iota\alpha ku)\rho-\alpha' up-\alpha
u'p-\alpha up'-\alpha\gamma^2up(VV'+uu')\}\nonumber\\
&&+c_2\{(-\iota\omega+\iota\alpha ku)p+\alpha'up+\alpha
u'p+\alpha up'+\alpha\gamma^2up(VV'+uu')\}\nonumber\\
&&+c_3(\rho+p)\left[-\iota\omega\gamma^2u +\iota
k\alpha(1+\gamma^2u^2)-\alpha\left\{(1-2\gamma^2u^2)(1+\gamma^2u^2)\frac{u'}{u}
-2\gamma^4u^2VV'\right\}\right]\nonumber\\
&&+c_4(\rho+p)[\gamma^2V(-\iota\omega +\iota k\alpha u)
+\alpha\gamma^2u\{(1+2\gamma^2V^2)V'+2\gamma^2uVu'\}]=0,\\
\label{4.13}
&&c_1\rho\gamma^2u\{(1+\gamma^2V^2)V'+\gamma^2uVu'\}
+c_2p\gamma^2u\{(1+\gamma^2V^2)V'+\gamma^2uVu'\}\nonumber\\
&&c_3(\rho+p)\gamma^2\left[\gamma^2uV\left(\frac{-\iota\omega}{\alpha}+\iota
ku\right)+\{(1+2\gamma^2V^2)(1+2\gamma^2u^2)-\gamma^2V^2\}V'
+2\gamma^2uVu'(1+2\gamma^2u^2)\right]\nonumber\\
&&+c_4(\rho+p)\gamma^2\left[(1+\gamma^2V^2)\left(\frac{-\iota\omega}{\alpha}+\iota
ku\right)+\gamma^2u\{(1+4\gamma^2u^2)uu'+4(1+\gamma^2V^2)VV'\}\right]=0,\\
\label{4.14}
&&c_1\rho\gamma^2\{a_z+uu'(1+\gamma^2u^2)+\gamma^2u^2VV'\}
+c_2[p\gamma^2\{a_z+uu'(1+\gamma^2u^2)+\gamma^2u^2VV'\}+p'+\iota kp]\nonumber\\
&&c_3(\rho+p)\gamma^2\left[(1+\gamma^2u^2)\left(\frac{-\iota\omega}{\alpha}+\iota
ku\right)+u'(1+\gamma^2u^2)(1+4\gamma^2u^2)
+2u\gamma^2\{(1+2\gamma^2u^2)VV'+a_z\}\right]\nonumber\\
&&+c_4(\rho+p)\gamma^4\left[uV\left(\frac{-\iota\omega}{\alpha}+\iota
ku\right)+u^2V'(1+4\gamma^2u^2)+2V\{(1+2\gamma^2u^2)uu'+a_z\}\right]=0,\\
\label{4.15}
&&c_1\left[\rho\gamma^2\left(\frac{-\iota\omega}{\alpha}+\iota
ku\right)+2\rho\gamma^2u\{a_z+\gamma^2(VV'+uu')\}+\rho\gamma^2u'+\rho'\gamma^2u\right]
+c_2\left[p\left\{\frac{-\iota\omega}{\alpha}(\gamma^2-1)+\iota
k\gamma^2u\right\}\right.\nonumber\\
&&\left.+2p\gamma^2u\{a_z+\gamma^2(VV'+uu')\}+p\gamma^2u'+p'\gamma^2u\right]
+c_3(\rho+p)\left\{2\gamma^4u\frac{-\iota\omega}{\alpha}+\gamma^2\iota
k(1+2\gamma^2u^2)\right.\nonumber\\
&&\left.-\gamma^2\frac{u'}{u}+6\gamma^6u^2(VV'+uu')
+\gamma^4(VV'+uu')+2\gamma^4uu'+\gamma^2a_z(1+2\gamma^2u^2)
\right\}\nonumber\\
&&+c_4(\rho+p)\left[2\gamma^4V\left(\frac{-\iota\omega}{\alpha}+\iota
ku\right)+2\gamma^4\{3\gamma^2uV(VV'+uu')+uV'+uVa_z)\}\right]=0.
\end{eqnarray}

\subsection*{Numerical Solutions}

We consider time lapse $\alpha=z$ and assume that $V=u$ in the
stiff fluid of constant density. Using mass conservation law in
three dimensions (with constant mass density), we obtain the value
of $u=\frac{1}{\sqrt{2+z^2}}$. Since the magnetic field is absent,
the plasma contains the longitudinal electron plasma waves, ion
plasma waves or the transverse electromagnetic waves.

The determinant of the coefficients of constants $c_1,~c_2,~c_3$ and
$c_4$ from Eqs.(\ref{4.12})-(\ref{4.15}) gives a complex dispersion
equation. The real part leads to dispersion equation of the type
$A_1(z)k^4+
A_2(z,\omega)k^3+A_3(z,\omega)k^2+A_4(z,\omega)k+A_5(z,\omega)=0.$
From the imaginary part, we obtain equation of the type
$A_1(z)k^3+A_2(z,\omega)k^2+A_3(z,\omega)k+A_4(z,\omega)=0.$ The
dispersion relation from the real part gives four values of $k$ out
of which two are interesting. The other two values turn out to be
imaginary in the whole region. The imaginary part of the determinant
gives three values out of which two are complex conjugates. The wave
numbers obtained from the real part are shown in Figures 3 and 4
while Figure 5 demonstrates the value of $k$ obtained from the
imaginary part.

\begin{figure}
\center \epsfig{file=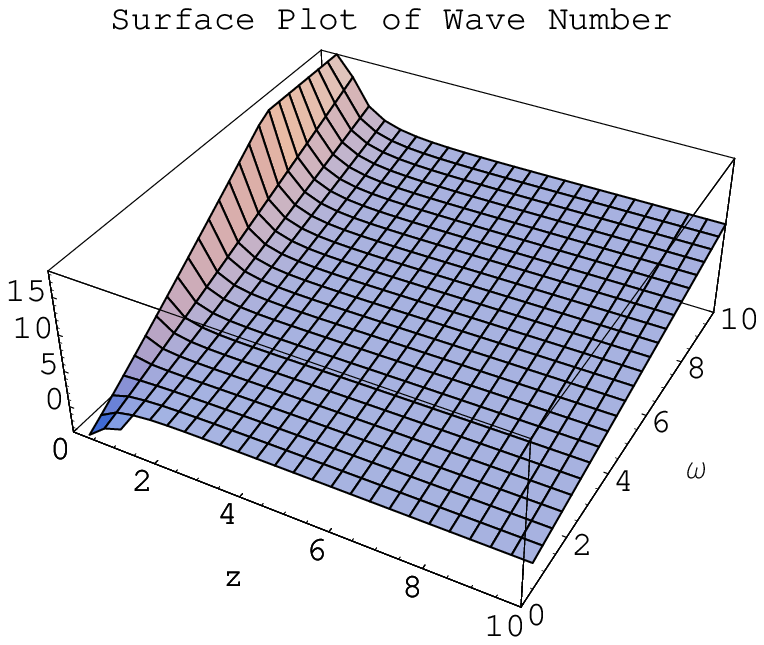,width=0.4\linewidth} \center
\epsfig{file=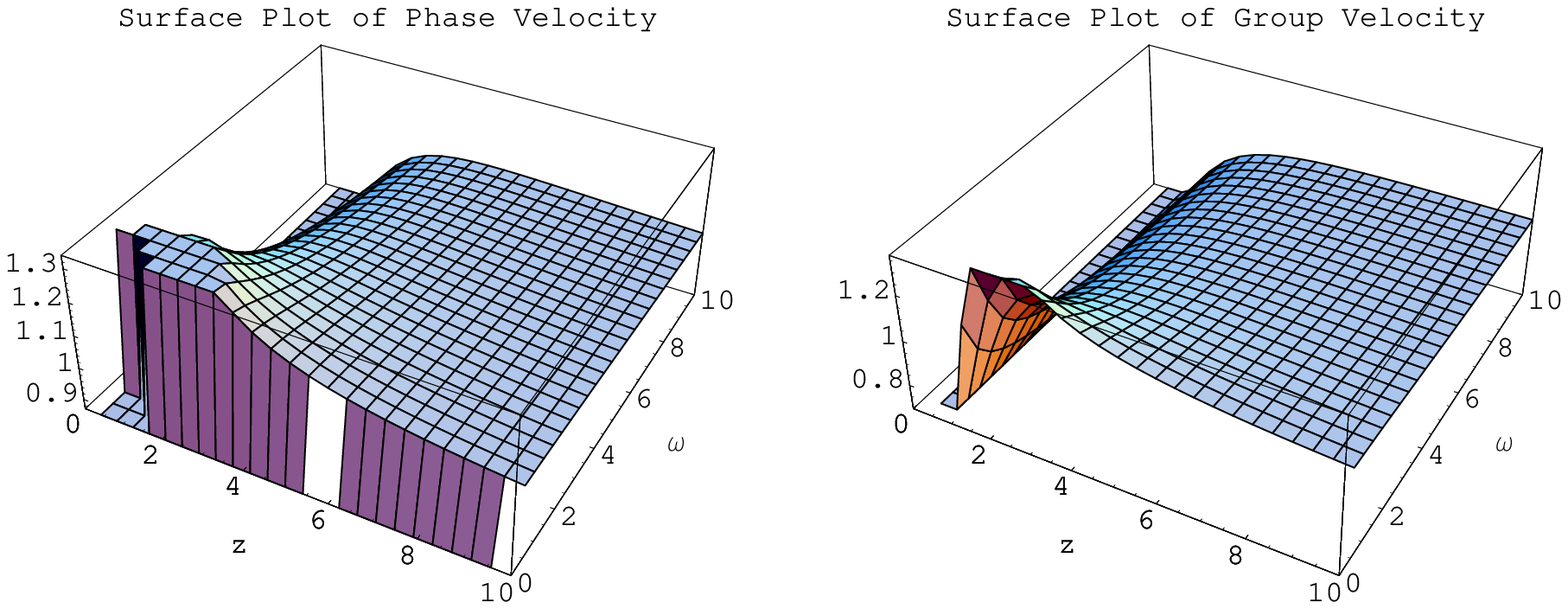,width=0.8\linewidth} \center
\epsfig{file=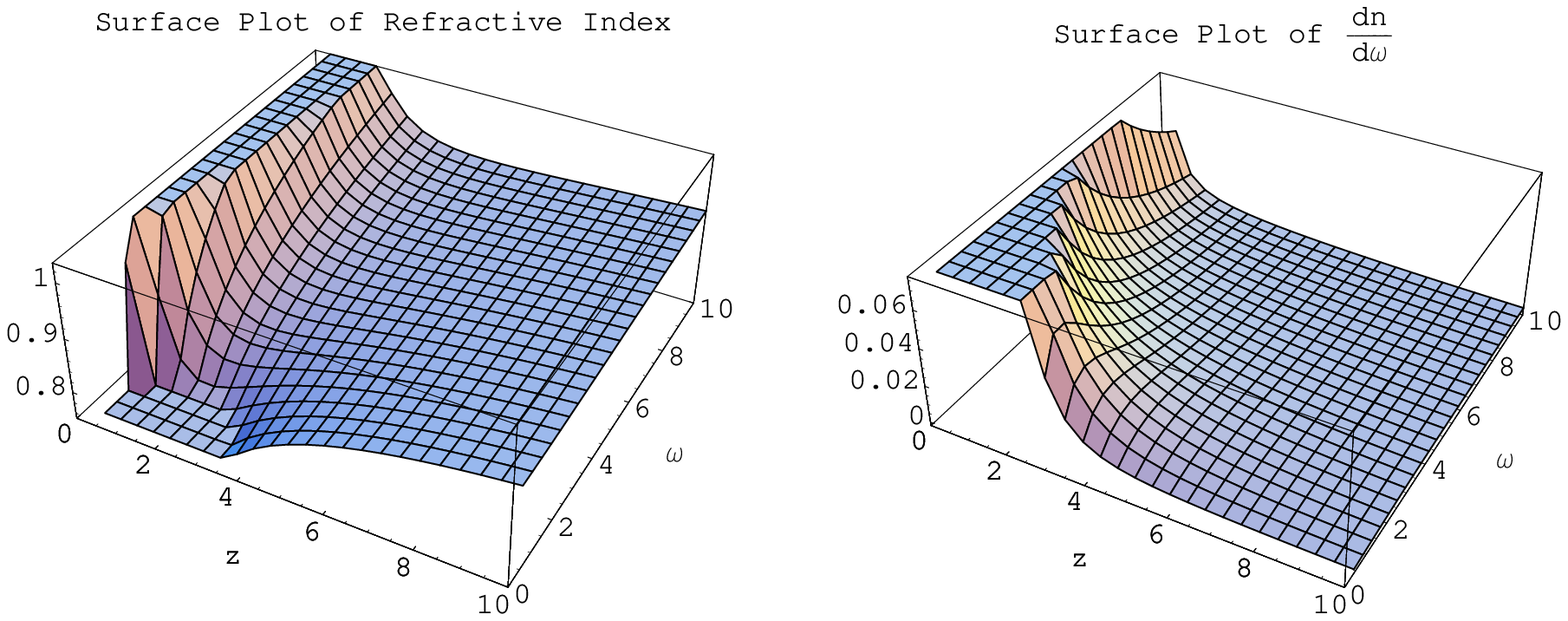,width=0.8\linewidth} \caption{The wave number
decreases when waves move away from the event horizon. A small
region is showing properties of Veselago medium. The dispersion is
not normal in most of the region.}
\end{figure}
We see from Figure 3 that $k$ is positive for
$0<z\leq10,~0.011<\omega\leq10$ while it is negative for
$0<z\leq1,~0<\omega\leq0.011$. It is interesting to note that in
this small region, the phase velocity and refractive index are
negative whereas the group velocity is positive. The fluid shows the
properties of Veselago medium \cite{Veselago}. The wave number
decreases with the increase in $z$ and increases with the increase
in angular frequency in the region $1<z\leq10,~0.011<\omega\leq10$.
In the neighborhood of $\omega=0$, the phase velocity is large. It
decreases with the increase in $z$ with the exception of the
interval where it has complex values. The group velocity shows the
similar behavior with the exception that it is complex at
$\omega=0$. In the region $0<z\leq2,~2<\omega\leq10$, the index of
refraction is greater than one and also $\frac{dn}{d\omega}>0$. The
dispersion is therefore normal in this small region but not normal
otherwise.

\begin{figure}
\center \epsfig{file=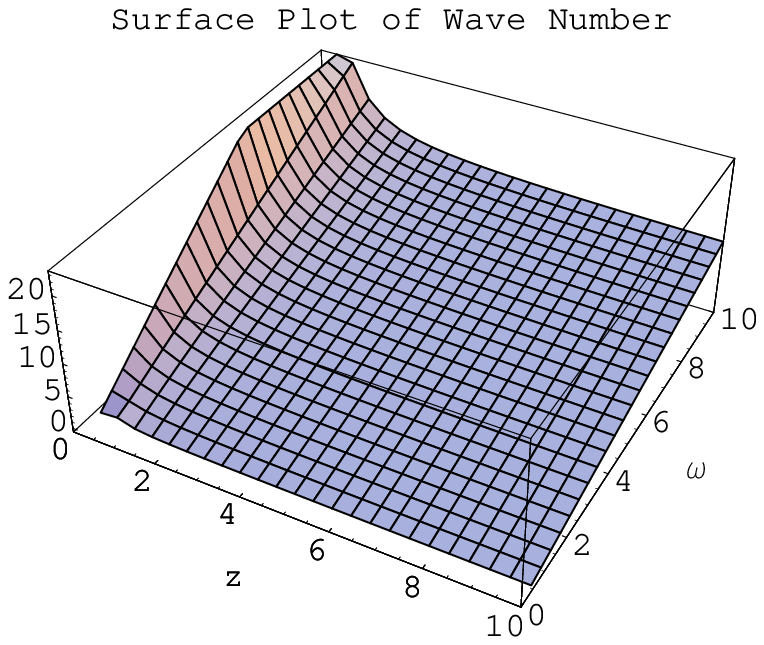,width=0.4\linewidth} \center
\epsfig{file=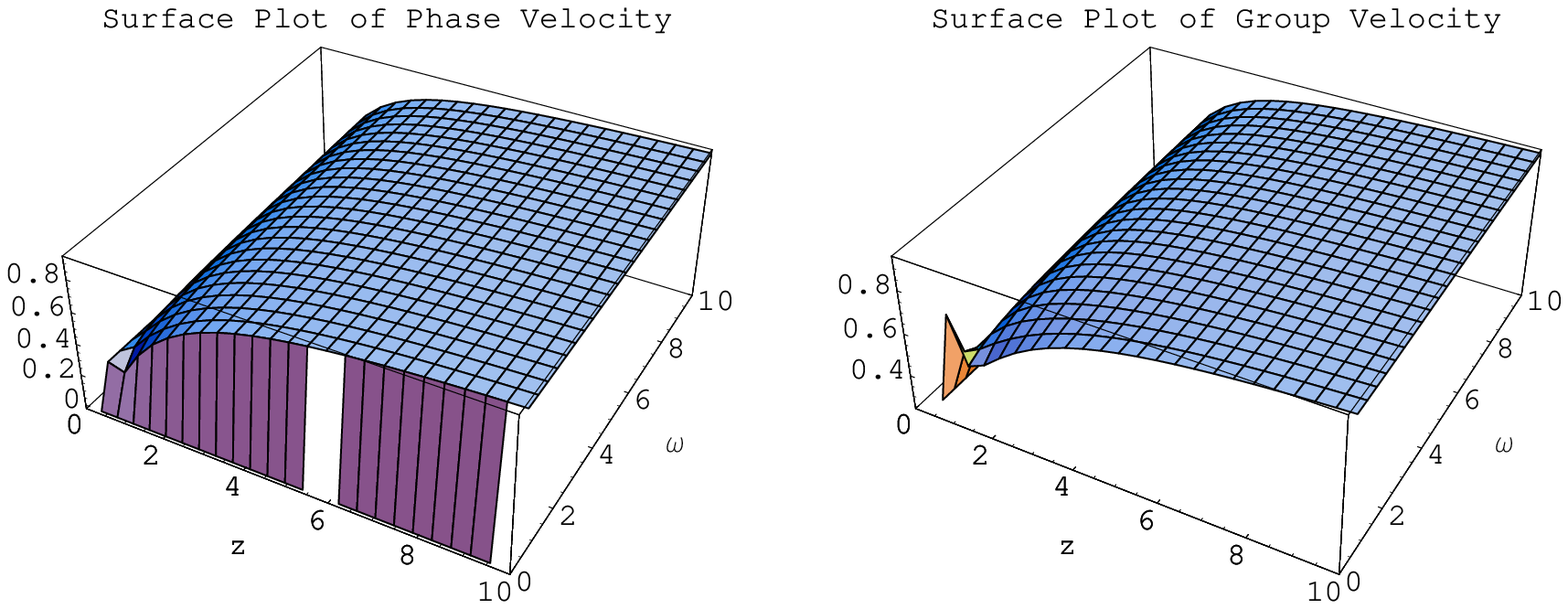,width=0.8\linewidth} \center
\epsfig{file=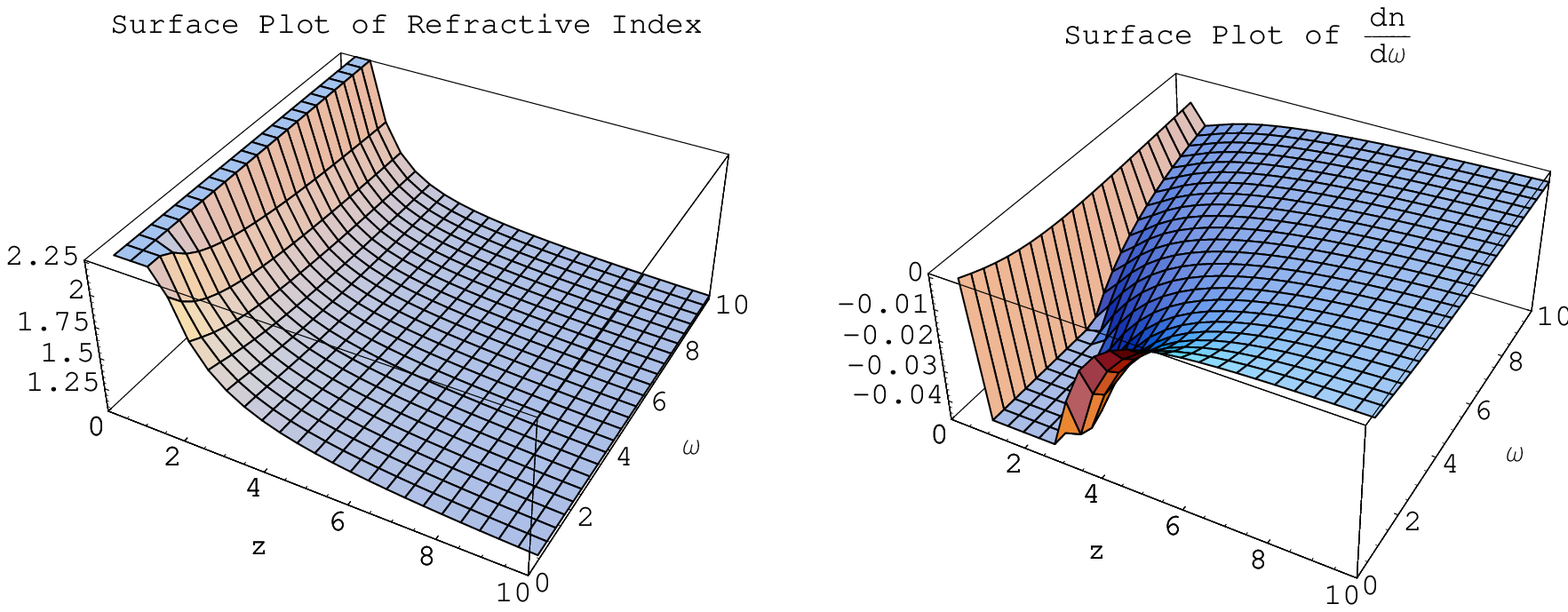,width=0.8\linewidth} \caption{The wave number
decreases as the waves move away from the event horizon. The
dispersion of waves is found to be not normal.}
\end{figure}
In Figure 4, the wave number becomes imaginary in the region
$0\leq\omega\leq0.5$ and hence evanescent waves exist there. The
wave number increases with an increase in $\omega$ and the waves
lose energy as we go away from the horizon. In the region
$0<z\leq10,~0.5<\omega\leq10$, the wave number as well as the phase
and group velocities are real. The refractive index is greater near
the horizon due to immense gravitational field and decreases away
from the horizon. The change of refractive index with respect to the
angular velocity is negative in the whole region. Thus the
dispersion is not normal.

\begin{figure} \center
\epsfig{file=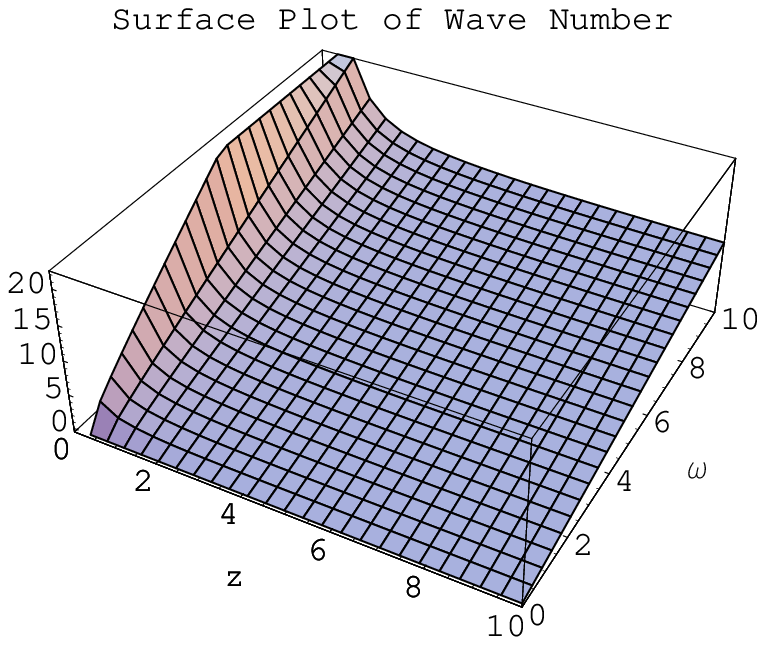,width=0.4\linewidth} \center
\epsfig{file=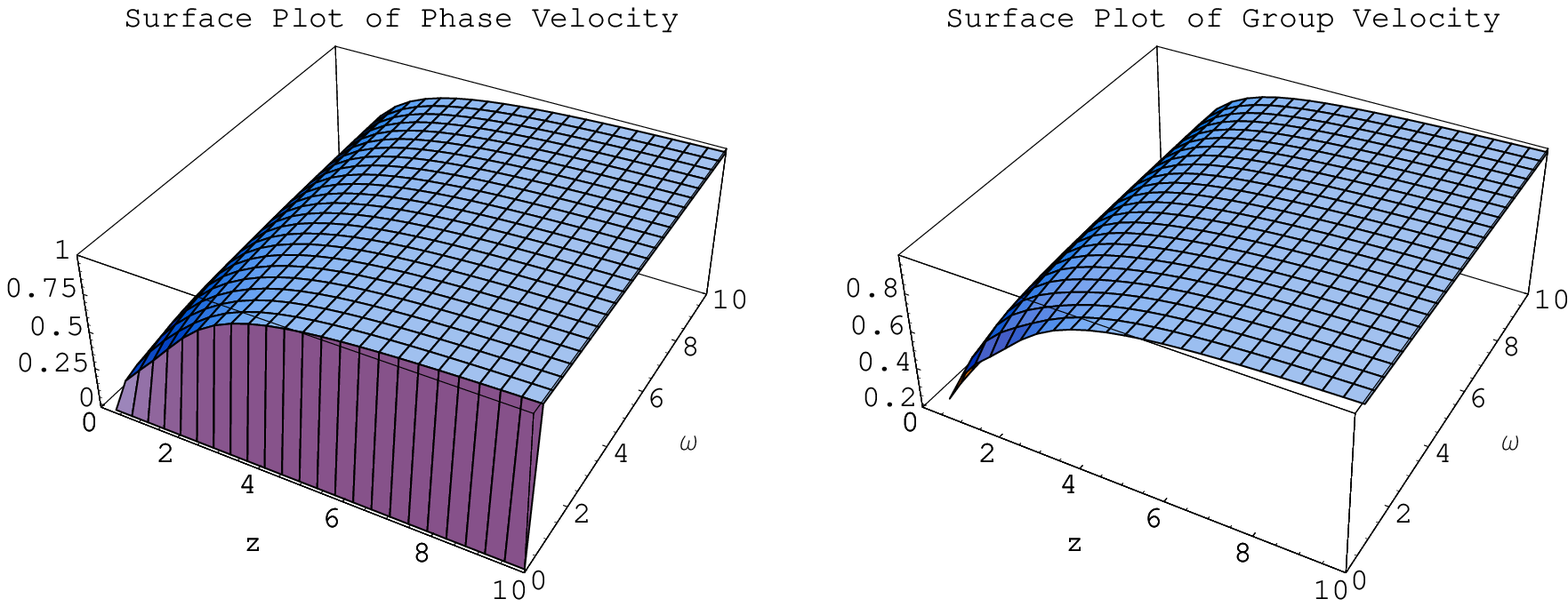,width=0.8\linewidth} \center
\epsfig{file=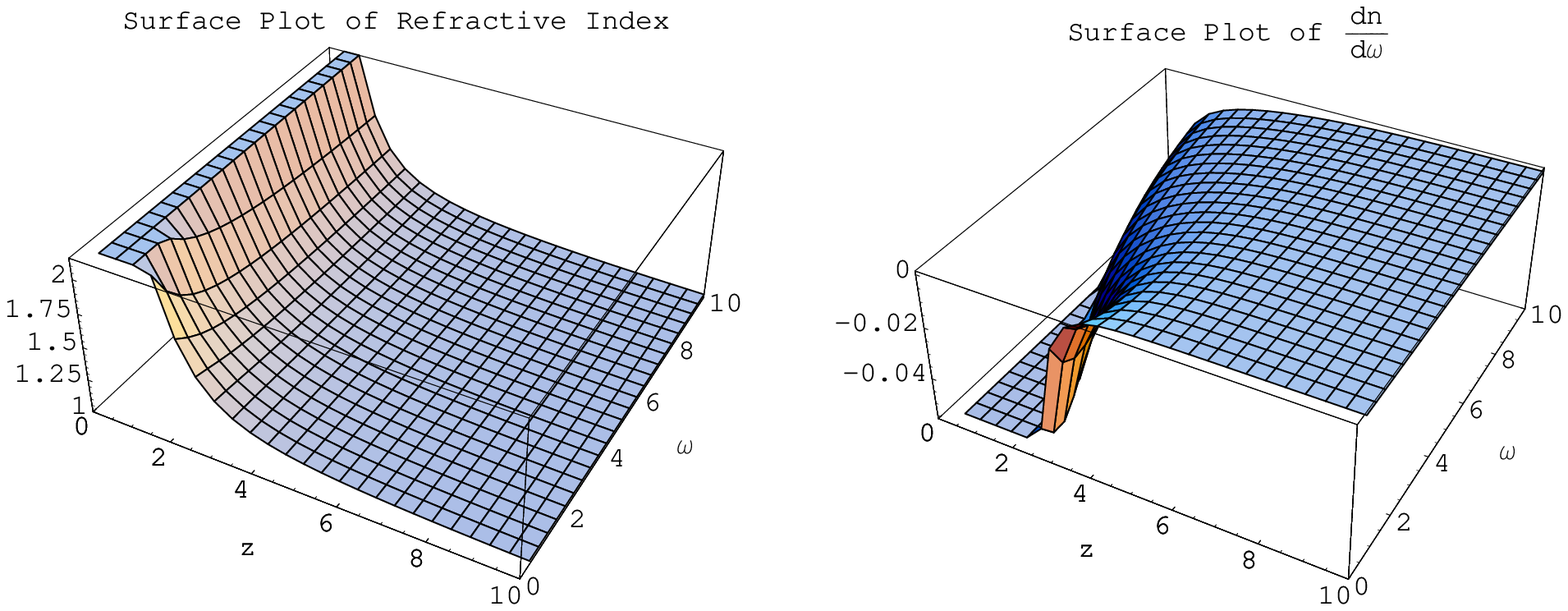,width=0.8\linewidth} \caption{The wave number
decreases as the waves move away from the horizon of the black
hole. The waves are not normally dispersed.}
\end{figure}
Figure 5 gives the description of the wave number obtained from the
imaginary part of the determinant. The wave number increases as they
move towards the horizon of the black hole. It also increases as the
angular frequency of waves increases. The waves are evanescent in
the region $0\leq z<0.112$ while the real waves lie in the region
$0.112\leq z\leq10$. The phase velocity is zero at $z=0.112$ and it
increases towards 1 as $z$ increases. The refractive index is
greater than one in this region but its change with respect to
angular frequency is negative which shows that dispersion is not
normal.

\section{Rotating Magnetized Background}

This is the most general case. The perturbed GRMHD equations for
this situation are given in Appendix C. Since these equations
include the additional magnetic field, therefore the Fourier
analysis procedure demands that the magnetic field perturbations
should have harmonic space and time dependence, i.e., of the form
$\textbf{b}\sim e^{-\iota(\omega t-k z)}$. These perturbed variables
admit the following notations
\begin{eqnarray}{\setcounter{equation}{1}}\label{5.1}
b_z(t,z)=c_5e^{-\iota(\omega t-kz)},\quad
b_x(t,z)=c_6e^{-\iota(\omega t-kz)}.
\end{eqnarray}

When we substitute these values in component form of perturbed
Eqs.(\ref{5.8}) and (\ref{5.9}) (Appendix C), it follows that
\begin{eqnarray*}
c_5\left(-\frac{\iota \omega}{\alpha}+\iota ku\right)=0,\quad\iota
kc_5=0.
\end{eqnarray*}
Both these equations yield that $c_5=0$ and hence $b_z=0$.
Substituting this value in addition to the values from
Eqs.(\ref{4.4}) and (\ref{5.1}) in Eqs.(\ref{5.7}),
(\ref{5.10})-(\ref{5.13}), we obtain the following Fourier analyzed
equations
\begin{eqnarray}
\label{5.14}
&&-c_3\{(\alpha \lambda)'+\iota
k\alpha\lambda\}+c_4(\alpha'+\iota
k \alpha)-c_6\{(\alpha u)'-\iota \omega +\iota k \alpha u \}=0,\\
\label{5.15} &&c_1\left\{\rho\left(-\iota \omega+\iota k \alpha
u\right)-\alpha'up-\alpha u'p-\alpha up'-\alpha
up\gamma^2(VV'+uu')\right\}\nonumber\\
&&+c_2\left\{p\left(-\iota \omega+\iota k\alpha
u\right)+\alpha'up+\alpha u'p+\alpha up'+\alpha
up\gamma^2(VV'+uu')\right\}\nonumber\\
&&+c_3(\rho+p)\left[-\iota\omega\gamma^2u+\iota
k\alpha(1+\gamma^2u^2)-\alpha\left\{(1-2\gamma^2u^2)
(1+\gamma^2u^2)\frac{u'}{u}-2\gamma^4u^2VV'\right\}\right]\nonumber\\
&&+c_4(\rho+p)\gamma^2\left[(-\iota\omega+\iota k\alpha u)V
+\alpha u\{(1+2\gamma^2V^2)V'+2\gamma^2uVu'\}\right]=0,\\
\label{5.16} &&c_1\rho\gamma^2u\{(1+\gamma^2V^2)V'+\gamma^2uVu'\}
+c_2p\gamma^2u\{(1+\gamma^2V^2)V'+\gamma^2uVu'\}\nonumber\\
&&+c_3\left[-\left\{(\rho+p)\gamma^4 u V-\frac{\lambda
B^2}{4\pi}\right\}\frac{\iota \omega}{\alpha}+\iota k
u\left\{(\rho+p)\gamma^4 u V+ \frac{\lambda
B^2}{4\pi}\right\}\right.\nonumber\\
&&\left.+(\rho+p)\gamma^2\left\{\{(1+2\gamma^2u^2)(1+2\gamma^2V^2)
-\gamma^2V^2\}V'+2\gamma^2(1+2\gamma^2
u^2)uVu'\right\}+\frac{B^2u}{4\pi\alpha}(\lambda
\alpha)'\right]\nonumber\\
&&+c_4\left[-\left\{(\rho+p)\gamma^2(1+\gamma^2 V^2)+
\frac{B^2}{4\pi}\right\}\frac{\iota \omega}{\alpha}+\iota
ku\left\{(\rho+p)\gamma^2(1+\gamma^2
V^2)-\frac{B^2}{4\pi}\right\}\right.\nonumber\\
&&\left.+(\rho+p)\gamma^4u\{(1+4\gamma^2V^2)uu'+4(1+\gamma^2V^2)VV'\}
-\frac{B^2u\alpha'}{4\pi\alpha}\right]\nonumber\\
&&-\frac{B^2}{4\pi}c_6\left\{(1-u^2)\iota
k+\frac{\alpha'}{\alpha}(1-u^2)-uu'\right\}=0,\\
\label{5.17} &&c_1\gamma^2\rho[a_z+u\{(1+\gamma^2
u^2)u'+\gamma^2VuV'\}]+c_2[\gamma^2p\{a_z+u\{(1+\gamma^2
u^2)u'+\gamma^2VuV'\}\}+\iota kp+p']\nonumber\\
&&+c_3\left[-\left\{(\rho+p)\gamma^2(1+\gamma^2
u^2)+\frac{\lambda^2 B^2}{4\pi}\right\}\frac{\iota
\omega}{\alpha}+\left\{(\rho+p)
\gamma^2(1+\gamma^2u^2)-\frac{\lambda^2 B^2}{4\pi}\right\}\iota
ku\right.\nonumber\\
&&\left.+\left\{(\rho+p)\gamma^2\{u'(1+\gamma^2 u^2)(1+4\gamma^2
u^2) +2u\gamma^2 \{(1+2\gamma^2 u^2)VV'+a_z\}\}-\frac{\lambda B^2
u}{4\pi\alpha}(\alpha \lambda)'\right\}\right]\nonumber\\
&&+c_4\left[-\left\{(\rho+p) \gamma^4 u V- \frac{\lambda
B^2}{4\pi}\right\}\frac{\iota
\omega}{\alpha}+\left\{(\rho+p)\gamma^4 u
V+\frac{\lambda B^2}{4\pi}\right\}\iota ku\right.\nonumber\\
&&\left.+\left\{(\rho+p)\gamma^4 \{u^2V'(1+4\gamma^2
V^2)+2V\{a_z+uu'(1+2\gamma^2 u^2)\}\}+\frac{\lambda B^2 \alpha'
u}{4\pi\alpha}\right\}\right]\nonumber\\
&&+\frac{B^2}{4 \pi}c_6\left\{\lambda (1-u^2)\iota
k+\lambda\frac{\alpha'}{\alpha}(1-u^2)-\lambda uu'+\frac{(\lambda
\alpha)'}{\alpha}\right\}=0,\\
\label{5.18}
&&c_1\gamma^2\left[\rho\left(\frac{-\iota\omega}{\alpha}+\iota
ku\right)+2\rho u\{a_z+\gamma^2(VV'+uu')\}+\rho u'+\rho'u\right]
+c_2\left[p\left\{\frac{-\iota\omega}{\alpha}(\gamma^2-1)+\iota
k\gamma^2u\right\}\right.\nonumber\\
&&\left.+2p\gamma^2u\{a_z+\gamma^2(VV'+uu')\}+p\gamma^2u'+p'\gamma^2u\right]
+c_3\left[(\rho+p)\left\{2\gamma^4u\frac{-\iota\omega}{\alpha}+\gamma^2\iota
k(1+2\gamma^2u^2)\right.\right.\nonumber\\
&&\left.\left.-\gamma^2\frac{u'}{u}+6\gamma^6u^2(VV'+uu')
+\gamma^4(VV'+uu')+2\gamma^4uu'+\gamma^2a_z(1+2\gamma^2u^2)\right\}
\right.\nonumber\\
&&\left.+\frac{B^2}{4\pi\alpha}\{\lambda(\alpha\lambda)'
-u(\alpha\lambda)'(u\lambda-V)-\iota
(k\alpha u+\omega)\lambda (u\lambda-V)\right]
+c_4\left[2\gamma^4(\rho+p)
\left\{V\left(\frac{-\iota\omega}{\alpha}
+\iota ku\right)\right.\right.\nonumber\\
&&\left.\left.+\{3\gamma^2uV(VV'+uu')+uV'+uVa_z)\}\right\}
+\frac{B^2}{4\pi\alpha}\{-(\alpha
\lambda)'+\alpha'u(u\lambda-V)+(\iota k \alpha
u+\iota \omega)(u\lambda-V)\}\right]\nonumber\\
&&+c_6\frac{B^2}{4\pi\alpha}\left[u(\alpha
\lambda)'+\{\alpha'-u(\alpha u)'+\iota
k(1-u^2)\}(u\lambda-V)\right]=0.
\end{eqnarray}

\subsection*{Numerical Solutions}

We take the same assumptions as in the case of rotating
non-magnetized plasma, i.e., $u=\frac{1}{\sqrt{z^2+2}}=V$.
Substituting this value in $V=\frac{1}{\alpha}+\lambda u$ of
\cite{Z1}, it turns out that $\lambda=1-\frac{\sqrt{z^2+2}}{z}$.
This shows that the magnetic field diverges near the horizon (as
discussed in \cite{Ha}). Also, $B^2$ is taken to be
$\frac{176}{7}$.

The matrix of the coefficients of constants $c_1,~c_2,~c_3,~c_4$ and
$c_6$ is of order $5\times5$ whose determinant gives a dispersion
equation quartic in $k$ from the real part and a quintic from the
imaginary part. The real part leads to four interesting solutions
whereas the imaginary part is difficult to solve to get exact
solutions. It is obvious that the magnetic field is non-zero and the
wave number is in arbitrary direction to the magnetic field. Since
the external magnetic field is present, the plasma is anisotropic.
This magnetic field has profound effects in plasma wave modes
discussed in the previous section.

The real part gives dispersion relation of the type $A_1(z)k^4+
A_2(z,\omega)k^3+A_3(z,\omega)k^2+A_4(z,\omega)k+A_5(z,\omega)=0.$
It gives four values of $k$ out of which two are complex conjugate
roots and are not interesting. The two interesting roots provided by
the real part are shown in Figures 6 and 7.
\begin{figure}
\center \epsfig{file=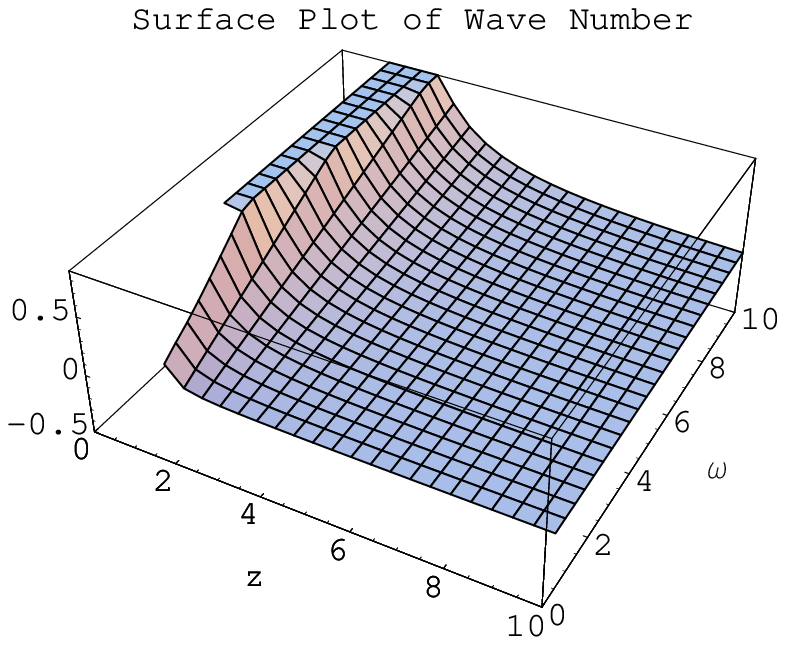,width=0.4\linewidth} \center
\epsfig{file=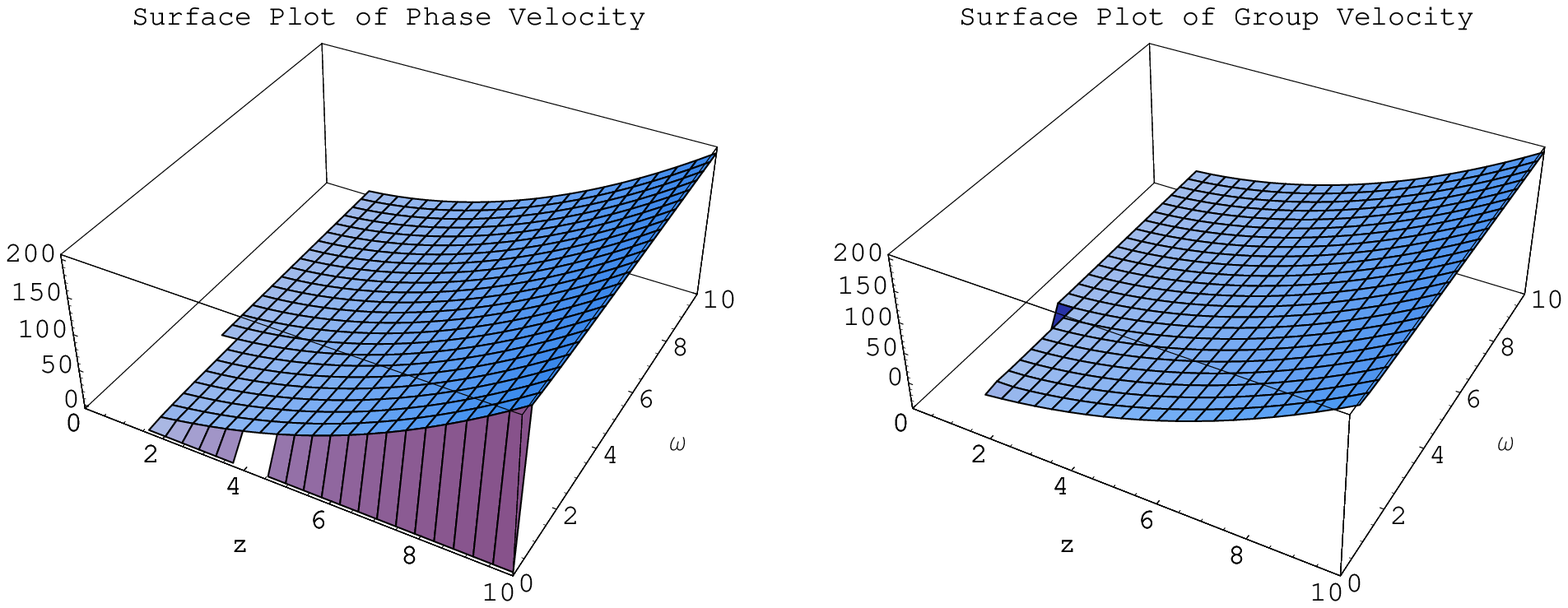,width=0.8\linewidth} \center
\epsfig{file=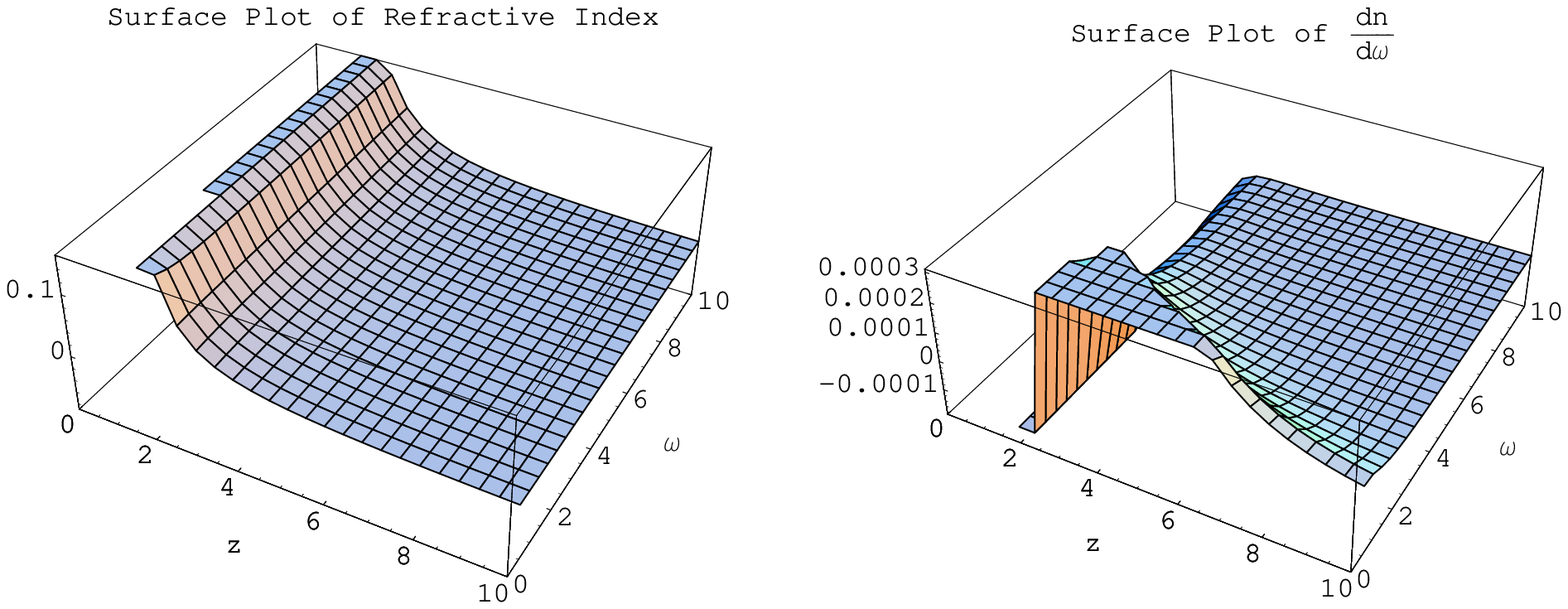,width=0.8\linewidth} \caption{The wave number
decreases as the waves move away from the event horizon. The phase
and group velocities are equal in most of the region. The
dispersion is not normal.}
\end{figure}

In Figure 6, waves are evanescent in the regions $0<z<1.65$ and
$0<\omega<2.5\times10^{-4}$. The wave number becomes infinite at
zero angular frequency as well as at $z=0$ and hence waves do not
exist there. Real waves are present in rest of the region. We see
that the wave number is large near the horizon and decreases
gradually afterwards as $z$ increases. Also, $k$ increases with an
increase in the angular frequency. The phase and group velocities
increase as we go away from the horizon. Also, the phase velocity is
zero at zero angular frequency but the group velocity is infinite
there. The refractive index is less than one in the whole region and
hence the dispersion is not normal.
\begin{figure}
\center \epsfig{file=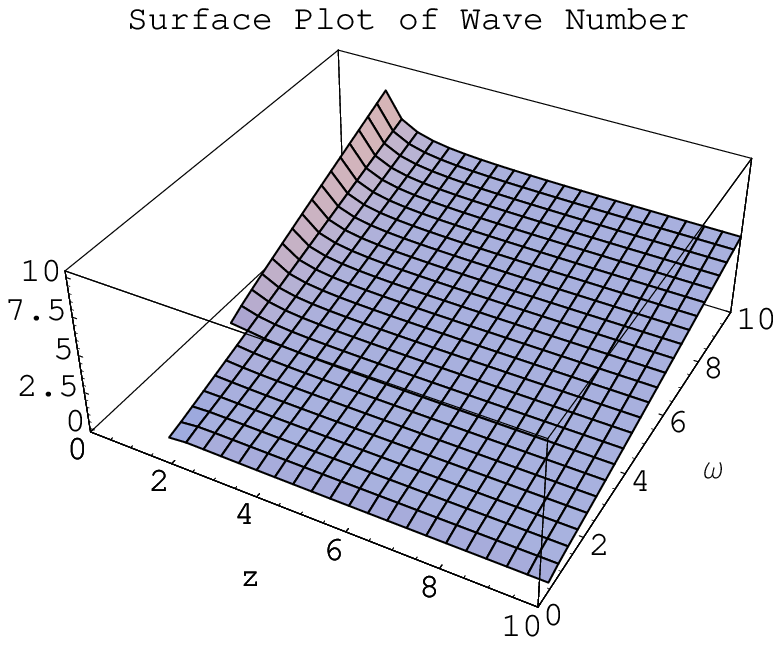,width=0.4\linewidth} \center
\epsfig{file=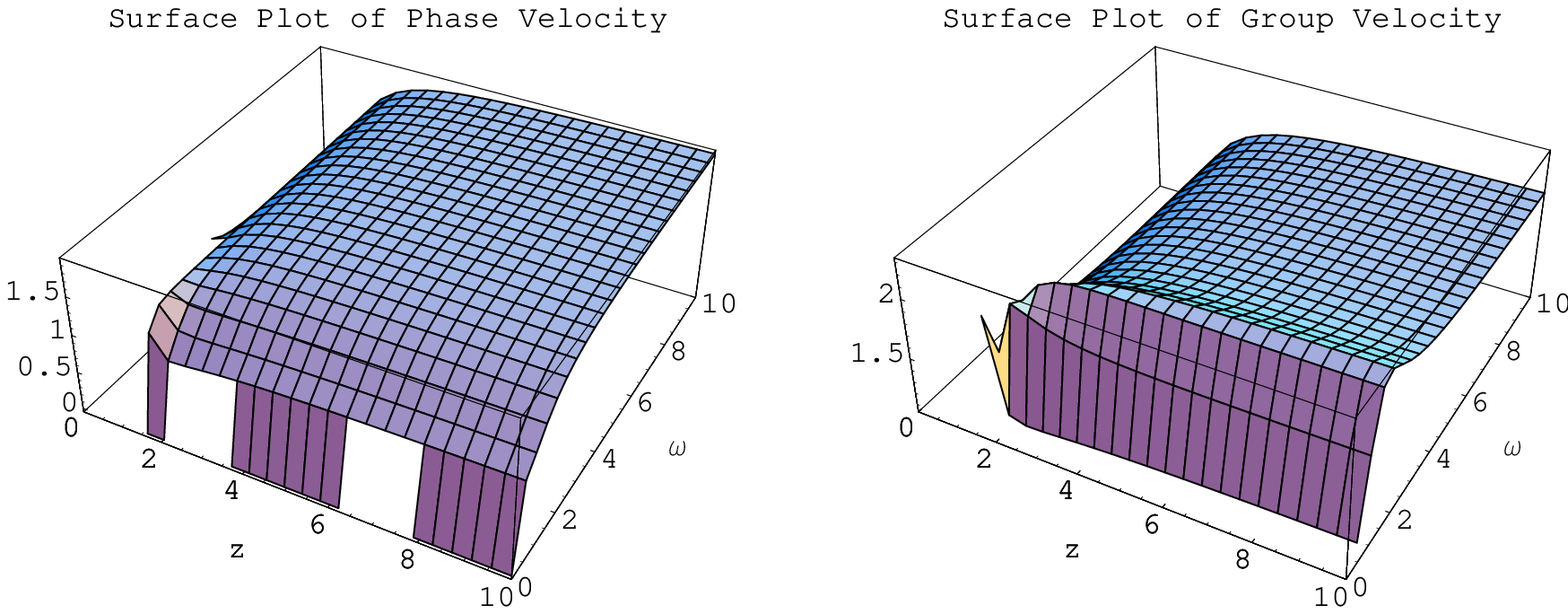,width=0.8\linewidth} \center
\epsfig{file=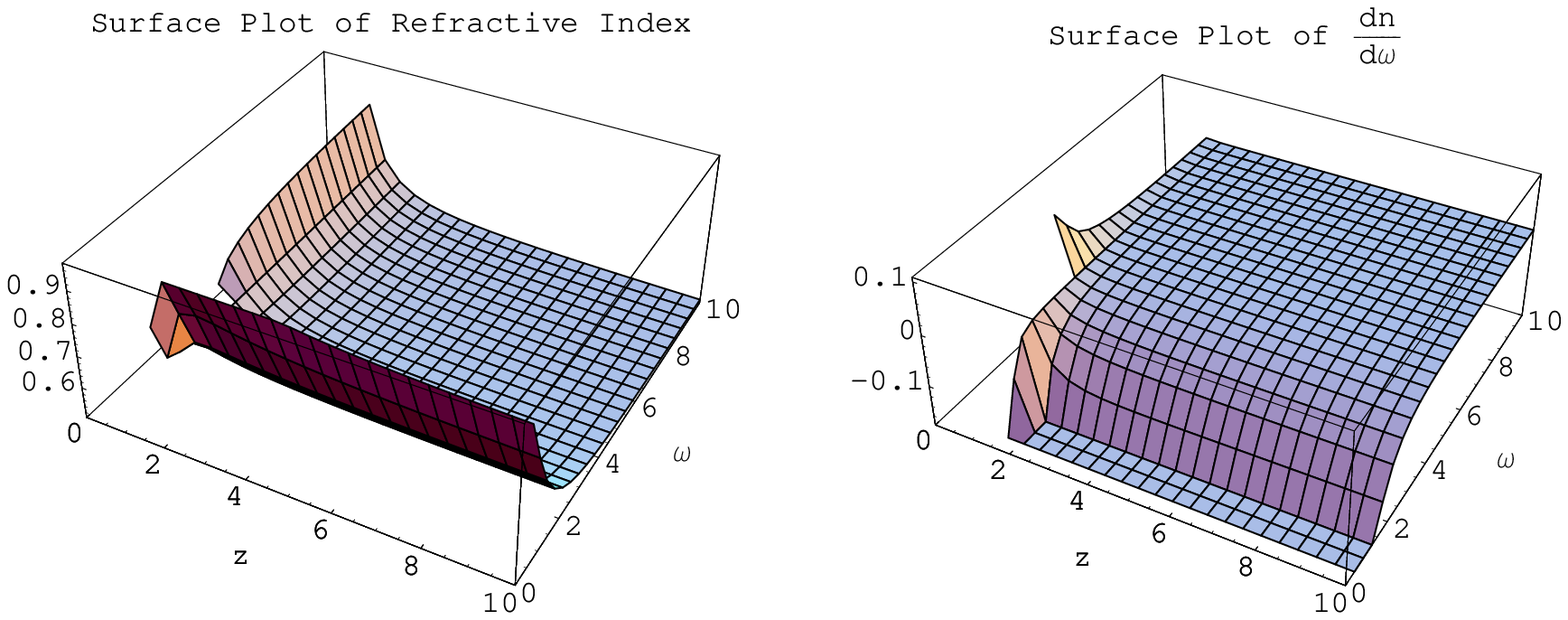,width=0.8\linewidth} \caption{The waves
decreases in number as we move away from the event horizon of the
black hole. The group velocity is greater than the phase velocity.
The dispersion is anomalous in the region.}
\end{figure}

Figure 7 shows that the evanescent waves lie in the regions
$0<z\leq1.65$ and $0\leq \omega \leq0.0005$. The wave number is
infinite at $z=0$ but decreases with an increase in $z$ and
increases with the increase in the angular frequency. The group
velocity is greater than the phase velocity which indicates that the
dispersion is anomalous. The value of refractive index is less than
one which confirms this indication.

The imaginary part of the determinant gives a dispersion relation
quintic in $k$, i.e.,
$A_1(z)k^5+A_2(z,\omega)k^4+A_3(z,\omega)k^3+A_4(z,\omega)k^2
+A_5(z,\omega)k+A_6(z,\omega)=0.$
This equation cannot be solved analytically for the exact solutions
and is solved numerically using software \emph{Mathematica}. The
roots are approximated in arrays for each point of the
two-dimensional mesh with equal step lengths $0.2$ for $z$ and
$\omega$. This equation gives one real root and two sets of complex
conjugate roots. The real data values for the root gives a real
interpolation function. This function represents the value of $k$
shown graphically which is used to investigate the properties of
medium.

When we substitute the assumptions, the approximated root becomes
infinite at $z=0$ due to which no wave exists there. We omit the
value $z=0$ and our mesh reduces to $0.2\leq
z\leq10,~0\leq\omega\leq 10$ for the interpolating function. The
corresponding real root is shown in Figure 8.
\begin{figure}
\center \epsfig{file=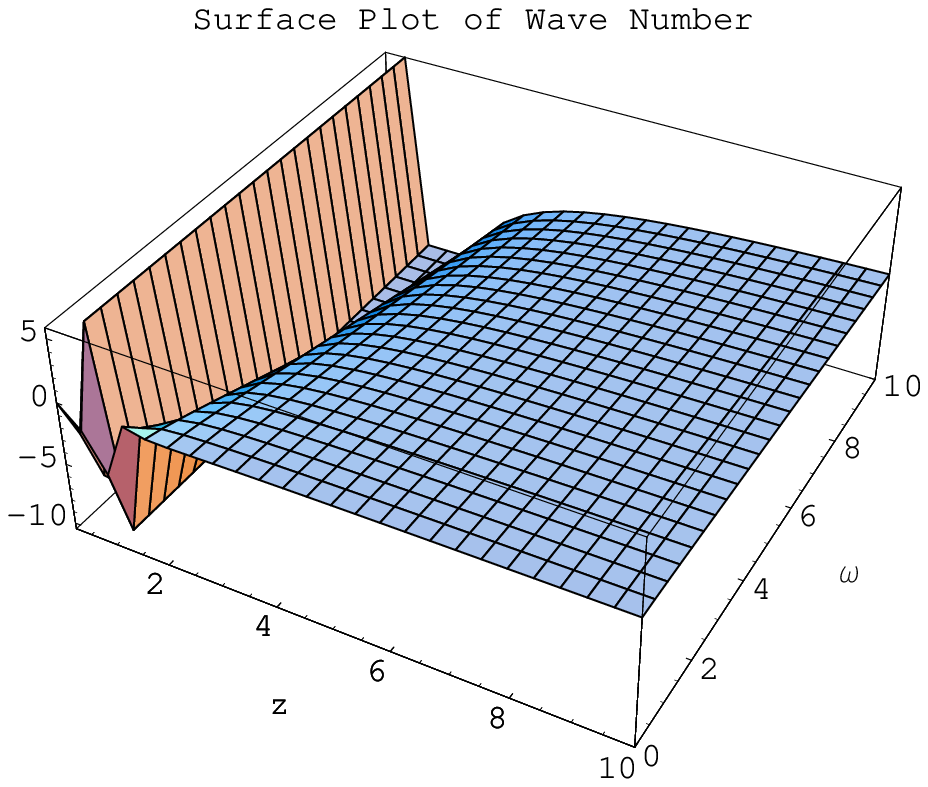,width=0.4\linewidth} \center
\epsfig{file=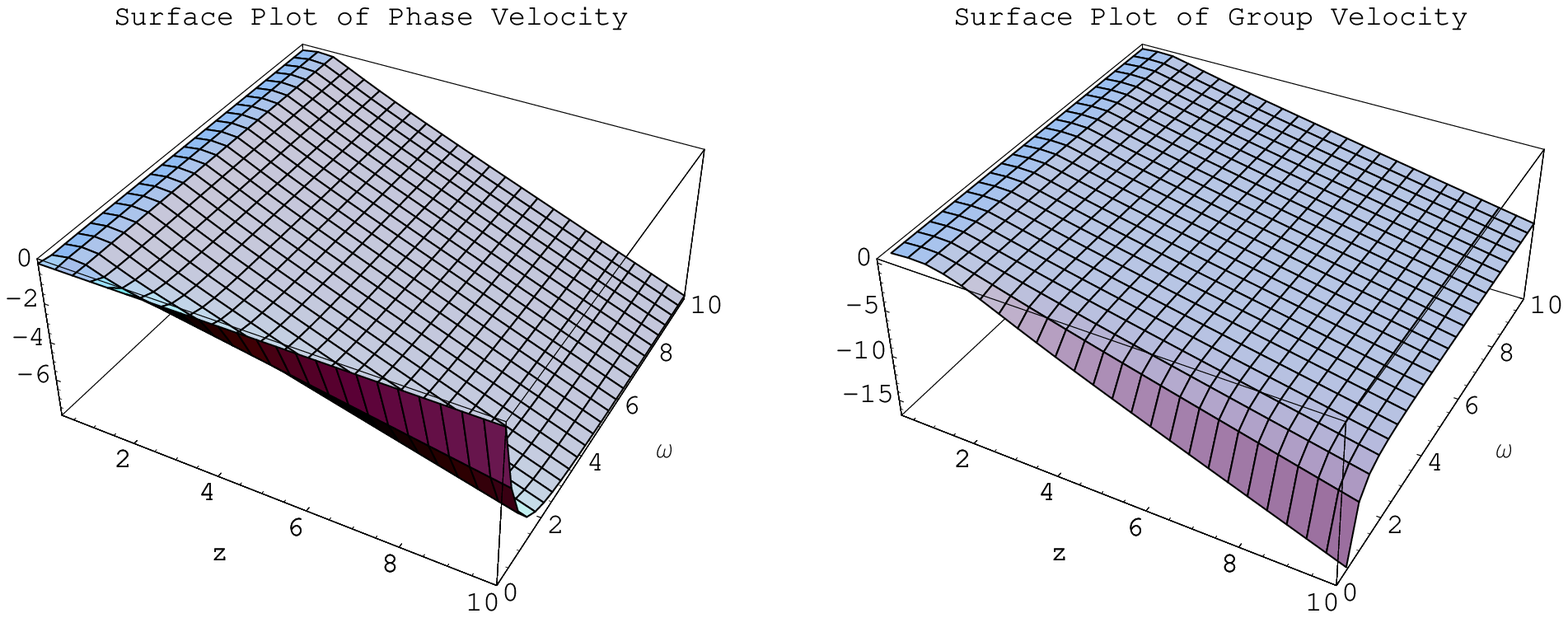,width=0.8\linewidth} \center
\epsfig{file=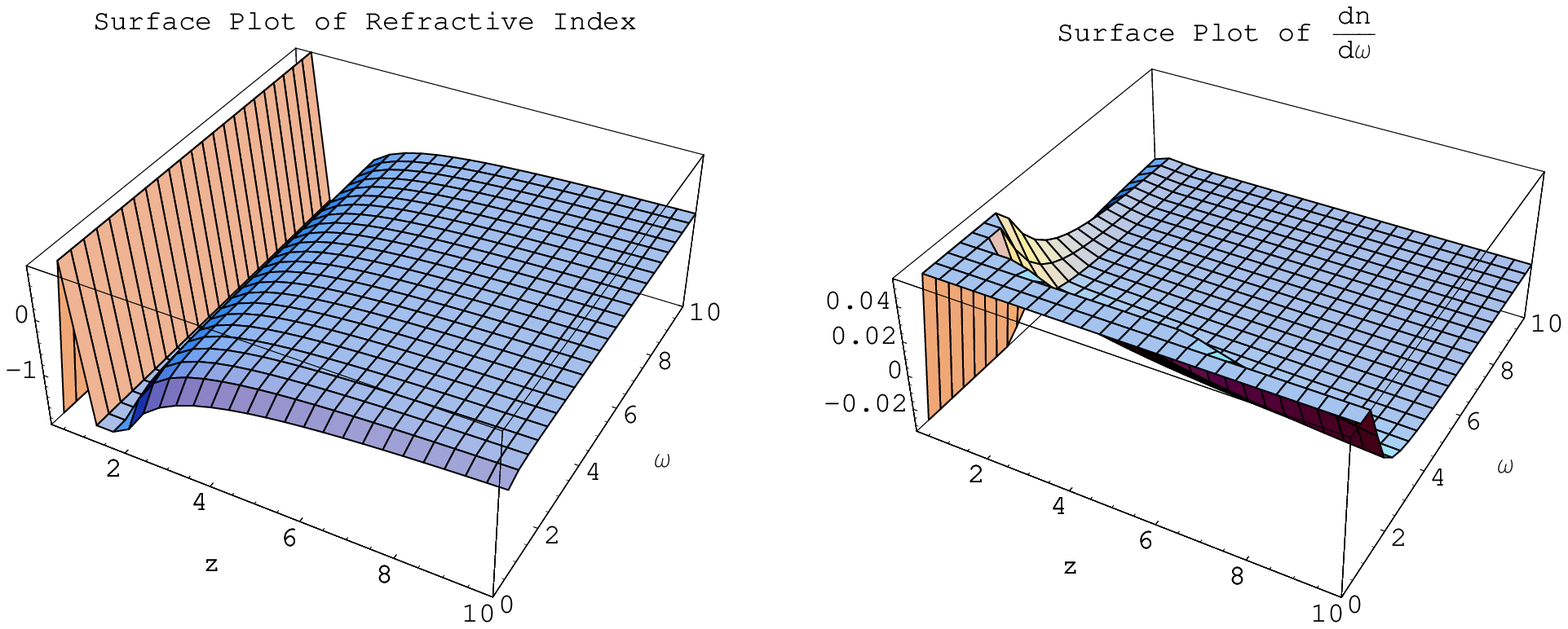,width=0.8\linewidth} \caption{Some of the
region possess the properties of Veselago medium. The phase
velocity is greater than the group velocity in most of the region
for which the dispersion is normal.}
\end{figure}

Figure 8 shows that the wave number is high near the horizon. It
abruptly decreases and gradually increases as we go away from the
horizon. The real wave number is shown in the figure. The wave
number becomes negative in the region $0.845\leq z\leq10$. Also, the
phase and group velocities as well as refractive index are negative
there which is the property of Veselago medium. The phase velocity
is greater than the group velocity in the whole region admitting
real waves. Also, the change in refractive index with respect to the
angular frequency is greater than zero for the region $1.61\leq
z\leq10,~0<\omega\leq10$, thus it admits normal dispersion of waves
\cite{WM}. The change in refractive index with respect to the
angular frequency is less than zero for the region $0.2\leq
z<1.61,~0<\omega\leq10$ which shows anomalous dispersion of waves.

\section{Conclusion}

We derive the GRMHD equations for the Schwarzschild black hole
magnetosphere in 3+1 formalism. We discuss one dimensional
perturbations in perfect MHD condition. This has been explored for
isothermal state of plasma. These equations are written in component
form and then Fourier analyzed by using the assumption of plane
waves. The determinant of the coefficients is solved for the
following three cases in the magnetosphere:
\begin{enumerate}
\item Non-Rotating Background (either non-magnetized or magnetized);
\item Rotating Non-Magnetized Background;
\item Rotating Magnetized Background.
\end{enumerate}
The non-rotating background is the pure Schwarzschild geometry
outside the event horizon. The rotating background is the restricted
Kerr geometry in the vicinity of the event horizon which admits a
variable lapse function with negligible rotation. The dispersion
relations and consequently, the wave numbers are obtained for these
backgrounds. The properties of plasma are inferred on the basis of
this number and the corresponding quantities are obtained in the
form of graphs.

We find that the gravity effect can be observed in the
perturbations both in the non-rotating and rotating backgrounds.
The values of $k$ only show the gravitational effects on the
smooth harmonic wave type perturbations. The rotation of the
background adds this effect to the perturbed quantities. It can
also be seen that the magnetic field perturbations are influenced
in the case of rotating background whereas there is no effect of
gravitation on the perturbed magnetic field in the non-rotating
background. This implies that rotation affects the magnetic field.
It is interesting to note that wave numbers decrease and the phase
and group velocities increase as we go away from the horizon in
each case.

Recently, we have discussed the properties of the Schwarzschild
magnetosphere for the cold plasma case \cite{MS} and some
interesting consequences have been obtained about the wave analysis.
In this paper, we have extended this analysis by adding the pressure
effect (isothermal plasma). It turns out that pressure has brought
changes in the case of non-rotating, non-magnetized background. In
the case of cold plasma, normal dispersion of the waves was found
for the non-rotating background whereas in the similar case of
isothermal plasma, the dispersion of waves is not normal. In the
case of rotating non-magnetized background one case of normally
dispersed waves was found whereas in isothermal plasma with constant
pressure, no such case is observed. This means that the background
pressure ceases normal dispersion of waves.

In the case of rotating magnetized background, there is one case of
normal dispersion and two of anomalous dispersion. The case with
normal dispersion of waves occurs when the plasma admits the
properties of Veselago medium whereas in cold plasma there were two
such cases where the plasma shows the properties of this medium.
Thus the cold plasma equation of state provides more chances to the
fluid to admit the properties of Veselago medium.

It is worth interesting to note that the wave number becomes
infinite and consequently waves vanish at the event horizon in all
the cases. This corresponds to the well-known fact that no
information can be extracted from a black hole. The negative phase
velocity propagation regions are found in the cases of rotating
background which are discussed by Mackay et al. in \cite{Mackay}
according to which rotation of a black hole is required for negative
phase velocity propagation. The same results were found in the case
of cold plasma. Thus it can be deduced that the rotation brings
negative phase velocity whether pressure is involved in the
surrounding plasma or not. Mackay et al. \cite{MLS} found that
negative phase velocity propagation was neither observed in
Schwarzschild-anti de Sitter spacetime, nor found outside the event
horizon of a Schwarzschild black hole due to zero rotation. This is
what we have obtained for non-rotating case. Ross et al. \cite{BR}
also investigated the negative phase velocity propagation phenomenon
for Kerr-Newmann and Kerr-Sen metrics particularly close to the
outer event horizon when the magnitude of the charge is large.

The complex solutions of the dispersion equation are not included
because it is felt that these solutions will prove to be of little
significance compared to those which have been discussed. The
dispersion relations are obtained using 3+1 ADM formalism and
contain the factor of acceleration (depend on lapse function and
equals to $-g$) which makes them different from the usual MHD
dispersion relations. We have investigated the waves propagating in
a plasma influenced by the gravitational field. These internal
gravity waves interrupt the MHD waves, therefore we have cases of
dispersion in each hypersurface. It is mentioned here that some of
the figures have patches missing which are due to the existence of
complex values there. Mathematica cannot plot complex numbers with
real numbers. These complex numbers are shown by the gaps in the
figures. It would be interesting to investigate this analysis for
the
Kerr spacetime. Presently, this work is in progress.\\
\par\noindent
\par\noindent
{\bf Acknowledgment}

\vspace{0.5cm}

We would like to thank the Higher Education Commission (HEC)
Islamabad, Pakistan for the financial support through the {\it
Indigenous PhD 5000 Fellowship Program Batch-II}.

\vspace{0.3cm}

\renewcommand{\theequation}{A\arabic{equation}}
\setcounter{equation}{0}
\section*{Appendix A}

This Appendix includes the details to reach at the perturbed form of
the GRMHD equations meant for the non-rotating background. The
component form of these equations is also given.

The perturbed flow of isothermal plasma shall be characterized by
its fluid density $\rho$, pressure $p$, velocity \textbf{V} and
magnetic field \textbf{B} (as measured by the FIDO). The first order
perturbations in the above mentioned quantities are denoted by
$\delta \rho,~\delta p,~\delta\textbf{V}$ and $\delta \textbf{B}$.
Consequently, the perturbed variables take the following form
\begin{eqnarray}\label{8}
\rho&=&\rho^0+\delta{\rho}=\rho^0+\rho \tilde{\rho},\nonumber\\
p&=&p^0+\delta{p}=p^0+p \tilde{p},\nonumber\\
\textbf{B}&=&\textbf{B}^0+\delta{\textbf{B}}=\textbf{B}^0+B\textbf{b},\nonumber\\
\textbf{V}&=&\textbf{V}^0+\delta{\textbf{V}}=\textbf{V}^0+\textbf{v},
\end{eqnarray}
where $\rho^0,~p^0,~\textbf{B}^0$ and $\textbf{V}^0$ are unperturbed
quantities. The waves can propagate in $z$-direction due to
gravitation with respect to time $t$, hence the perturbed quantities
depend on $z$ and $t$.

For the non-rotating background, the perturbed flow of fluid is only
along $z$-axis. Thus the FIDOs measured four-velocity and magnetic
field are along $z$-axis which can be expressed as
$\textbf{V}=u(z)\textbf{e}_\textbf{z},~\textbf{B}=B(z)\textbf{e}_\textbf{z}$
and the Lorentz factor is $\gamma=\frac{1}{\sqrt{1-u^2}}$. The
perturbed quantities admit the following notations
\begin{eqnarray*}
\tilde{\rho}\equiv\frac{\delta\rho}{\rho}=\tilde{\rho}(t,z),\quad
\tilde{p}\equiv\frac{\delta p}{p}=\tilde{p}(t,z),\quad
\textbf{v}\equiv \delta
\textbf{V}=v_z(t,z)\textbf{e}_\textbf{z},\quad\textbf{b}\equiv
\frac{\delta \textbf{B}}{B} =b_z(t,z)\textbf{e}_\textbf{z}.
\end{eqnarray*}

Introducing perturbations given by Eq.(\ref{8}) in the GRMHD
Eqs.(\ref{3})-(\ref{7}), we obtain
\begin{eqnarray}
\label{3.2} &&\frac{\partial(\delta \textbf{B})}{\partial t}=\nabla
\times(\alpha\textbf{v}\times\textbf{B})
+\nabla\times(\alpha\textbf{V}\times\delta\textbf{B}),\\\label{3.3}
&&\nabla.(\delta\textbf{B})=0,\\\label{3.4}
&&\frac{\partial(\delta\rho+\delta p)}{\partial
t}+(\alpha\textbf{V}.\nabla)(\delta\rho+\delta
p)+(\rho+p)\gamma^2\textbf{V}.\frac{\partial\textbf{v}}{\partial
t}\nonumber\\
&&-\alpha(\rho+p)\textbf{v}.\nabla \ln u
+\alpha(\rho+p)(\nabla.\textbf{v})+(\delta\rho+\delta
p)\nabla.(\alpha\textbf{V})+(\delta\rho+\delta p)\gamma^2
\textbf{V}.(\alpha\textbf{V}.\nabla)\textbf{V}\nonumber\\
&&+2(\rho+p)\gamma^2(\textbf{V}.\textbf{v})(\alpha\textbf{V}.\nabla)
\ln\gamma+(\rho+p)\gamma^2(\alpha\textbf{V}.\nabla\textbf{V}).\textbf{v}
+(\rho+p)\gamma^2\textbf{V}.(\alpha\textbf{V}.\nabla)\textbf{v}=0,\\\label{3.5}
&&\{((\rho+p)\gamma^2+\frac{\textbf{B}^2}{4\pi})\delta_{ij}+(\rho+p)\gamma^4
V_iV_j-\frac{1}{4\pi}B_iB_j\}\frac{1}{\alpha}\frac{\partial
v^j}{\partial
t}+\frac{1}{4\pi}[\textbf{B}\times\{\textbf{V}\times\frac{1}{\alpha}
\frac{\partial(\delta\textbf{B})}{\partial t}\}]_i\nonumber\\
&&+(\rho+p)\gamma^2v_{i,j}V^j+(\rho+p)\gamma^4V_iv_{j,k}V^jV^k-\frac{1}{4\pi\alpha}
\{(\alpha\delta B_i)_{,j}-(\alpha\delta B_j)_{,i}\}B^j\nonumber\\
&&=-\gamma^2\{(\delta\rho+\delta
p)+2(\rho+p)\gamma^2(\textbf{V}.\textbf{v})\}a_i-(\delta
p)_{,i}+\frac{1}{4\pi\alpha}\{(\alpha B_i)_{,j}-(\alpha B_j)_{,i}\}\delta B^j \nonumber\\
&&-(\rho+p)\gamma^4
(v_iV^j+v^jV_i)V_{k,j}V^k-\gamma^2\{(\delta\rho+\delta p)V^j
+2(\rho+p)\gamma^2(\textbf{V}.\textbf{v})V^j+(\rho+p)v^j\}V_{i,j}\nonumber\\
&&-\gamma^4V_i\{(\delta\rho+\delta
p)V^j+4(\rho+p)\gamma^2(\textbf{V}.\textbf{v})V^j+(\rho+p)v^j\}V_{j,k}V^k,\\
\label{3.6}
&&\gamma^2\frac{1}{\alpha}\frac{\partial (\delta\rho+\delta
p)}{\partial
t}+\frac{2}{\alpha}(\rho+p)\gamma^4\textbf{V}.\frac{\partial\textbf{v}}{\partial
t}-2(\rho+p)\gamma^4(\textbf{V}.\textbf{v})(\textbf{V}.\nabla)\ln u
+6(\rho+p)\gamma^6(\textbf{V}.\textbf{v})\{\textbf{V}.(\textbf{V}.\nabla)\textbf{V}\}
\nonumber\\
&&+(\rho+p)\gamma^4\textbf{V}.(\textbf{v}.\nabla)\textbf{V}+2(\rho+p)\gamma^4
\textbf{V}.(\textbf{V}.\nabla)\textbf{v}+2(\delta\rho+\delta
p)\gamma^2\textbf{V}.\textbf{a}+2(\rho+p)\gamma^4(\textbf{V}.\textbf{v})
(\textbf{V}.\textbf{a})\nonumber\\
&&+(\rho+p)\gamma^2(\nabla.\textbf{v})+(\rho+p)\gamma^2\textbf{v}.\textbf{a}
+\gamma^2(\textbf{V}.\nabla)(\delta\rho+\delta p)
-\frac{1}{\alpha}\frac{\partial(\delta p)}{\partial t}
-(\rho+p)\gamma^2 (\textbf{v}.\nabla)\ln u\nonumber\\
&&+2(\delta \rho+\delta p)\gamma^4
\textbf{V}.(\textbf{V}.\nabla)\textbf{V}+(\delta\rho+\delta
p)\gamma^2(\nabla.\textbf{V})
+2(\rho+p)\gamma^4(\textbf{V}.\textbf{v})(\nabla.\textbf{V})
+2(\rho+p)\gamma^4\textbf{v}.(\textbf{V}.\nabla)\textbf{V}\nonumber\\
&&+\frac{1}{4\pi\alpha}[(\textbf{V}\times\textbf{B})
.(\nabla\times(\alpha\delta\textbf{B}))+
(\textbf{v}\times\textbf{B}).(\nabla\times(\alpha\textbf{B}))+(\textbf{V}
\times\delta\textbf{B}).(\nabla \times (\alpha\textbf{B}))\nonumber\\
&&+(\textbf{V}\times \textbf{B}).(\textbf{V}\times \frac{\partial
\delta\textbf{B}}{\partial t})
+(\textbf{V}\times\textbf{B}).(\frac{\partial\textbf{v}}{\partial
t}\times\textbf{B})]=0.
\end{eqnarray}
The component form of Eqs.(\ref{3.2})-(\ref{3.6}) is given as
follows
\begin{eqnarray}\label{3.7}
&&\frac{1}{\alpha}\frac{\partial b_z}{\partial t}=0,\\
\label{3.8}
&&b_{z,z}=0,\\
\label{3.9} &&\rho\frac{\partial \tilde{\rho}}{\partial
t}+u\alpha(\rho\tilde{\rho}_{,z}+p\tilde{p}_{,z})+p\frac{\partial
\tilde{p}}{\partial t}+\gamma^2u(\rho+p)\frac{\partial v_z}{\partial
t}+\alpha(\rho+p)(1+\gamma^2
u^2)v_{z,z}\nonumber\\
&&-(\tilde{\rho}-\tilde{p})\{(\alpha u p)'+\alpha \gamma^2 u^2 p
u'\} =\alpha(\rho+p)(1-2\gamma^2u^2)(1+\gamma^2 u^2)\frac{u'}{u}v_z,
\end{eqnarray}
\begin{eqnarray}
\label{3.10} &&(\rho+p)\gamma^2(1+\gamma^2
u^2)\left\{\frac{1}{\alpha} \frac{\partial v_z}{\partial t}+u
v_{z,z}\right\}
=-(\rho\tilde{\rho}+p\tilde{p})\gamma^2\{a_z+u(1+\gamma^2u^2)u'\}\nonumber\\
&&-\gamma^2(\rho+p)\{u'(1+\gamma^2 u^2)(1+4\gamma^2 u^2)
+2u\gamma^2a_z)\}v_z-p\tilde{p}'-\tilde{p}p',\\
\label{3.11} &&\rho\gamma^2\frac{1}{\alpha}\frac{\partial}{\partial
t} \tilde{\rho}+p\gamma^2\frac{1}{\alpha}\frac{\partial}{\partial
t}\tilde{p}+\frac{2}{\alpha}(\rho+p)\gamma^4u\frac{\partial
v_z}{\partial t}-p\frac{1}{\alpha}\frac{\partial\tilde{p}}{\partial
t}
+(\rho\tilde{\rho}+p\tilde{p})\gamma^2\{2ua_z+2\gamma^2u^2u'+u'\}\nonumber\\
&&+\gamma^2u(\rho_{,z}\tilde{\rho}
+\tilde{\rho}_{,z}\rho+p_{,z}\tilde{p}+\tilde{p}_{,z}p)
+(\rho+p)\gamma^2(1+2\gamma^2u^2)v_{z,z}\nonumber\\
&&+(\rho+p)\gamma^2\left\{(3\gamma^2uu'+a_z)(1+2\gamma^2 u^2)
-\frac{u'}{u}\right\}v_z=0.
\end{eqnarray}
The conservation law of rest-mass \cite{Z1} in three-dimensional
hypersurface for isothermal state of plasma $\alpha(\rho+p)\gamma
u=constant$ is used to obtain Eqs.(\ref{3.6}) and (\ref{3.9}).

\renewcommand{\theequation}{B\arabic{equation}}
{\setcounter{equation}{0}}
\section*{Appendix B}

When we consider non-magnetized isothermal plasma in rotating
background, i.e., $\textbf{B}=\textbf{0}$, the GRMHD Eqs.(\ref{3})
and (\ref{4}) vanish and Eqs.(\ref{5})-(\ref{7}) change into general
relativistic hydrodynamical (GRHD) equations given as follows
\begin{eqnarray}
\label{4.1} &&\frac{\partial (\rho+p)}{\partial t}+(\alpha
\textbf{V}.\nabla)(\rho+p)
+(\rho+p)\gamma^2\textbf{V}.\frac{\partial \textbf{V}}{\partial
t}+(\rho+p)\gamma^2\textbf{V}.(\alpha
\textbf{V}.\nabla)\textbf{V}+(\rho+p) \nabla.(\alpha\textbf{V})=0,\\
\label{4.2}
&&(\rho+p)\{\gamma^2\delta_{ij}+\gamma^4V_iV_j)\}\left(\frac{1}
{\alpha}\frac{\partial}{\partial
t}+\textbf{V}.\nabla \right)V^j=-(\rho+p)\gamma^2a_i-p_{,i},\\
\label{4.3} &&\gamma^2(\frac{1}{\alpha}\frac{\partial}{\partial
t}+\textbf{V}.\nabla)(\rho+p)
+2(\rho+p)\gamma^4\textbf{V}.(\frac{1}{\alpha}\frac{\partial}{\partial
t}+\textbf{V}.\nabla)\textbf{V}+2(\rho+p)\gamma^2\textbf{V}.\textbf{a}
-\frac{1}{\alpha}\frac{\partial p}{\partial t}\nonumber\\
&&+(\rho+p)\gamma^2\nabla.\textbf{V}=0.
\end{eqnarray}

We assume that the fluid is moving in $xz$-plane hence the fluid
measured four-velocity by FIDO can be written as
$\textbf{V}=V(z)\textbf{e}_\textbf{x}+u(z)\textbf{e}_\textbf{z}$ for
which the Lorentz factor becomes
$\gamma=\frac{1}{\sqrt{1-u^2-V^2}}.$ The following notations will be
used for the perturbed quantities
\begin{eqnarray*}
\tilde{\rho}\equiv\frac{\delta \rho}{\rho}=\tilde{\rho}(t,z),\quad
\tilde{p}\equiv\frac{\delta p}{p}=\tilde{p}(t,z),\quad
\textbf{v}\equiv \delta \textbf{V}=v_x(t,z)\textbf{e}_\textbf{x}+v
_z(t,z)\textbf{e}_\textbf{z}.
\end{eqnarray*}

Introducing perturbations in Eqs.(\ref{4.1})-(\ref{4.3}), it follows
that
\begin{eqnarray}\label{4.5}
&&\frac{\partial(\delta \rho+\delta p)}{\partial
t}+(\rho+p)\gamma^2\textbf{V}.\frac{\partial\textbf{v}}{\partial
t}+\alpha(\rho+p)\gamma^2\textbf{V}.(\textbf{V}.\nabla)\textbf{v}
+\alpha(\rho+p)(\nabla.\textbf{v})\nonumber\\
&&+(\delta \rho+\delta p)\nabla.(\alpha\textbf{V})+(\delta
\rho+\delta p)\gamma^2\textbf{V}.
(\alpha\textbf{V}.\nabla)\textbf{V}+(\alpha\textbf{V}.\nabla)(\delta
\rho+\delta p)\nonumber\\
&&+2\alpha(\rho+p)\gamma^2(\textbf{V}.\textbf{v})
(\textbf{V}.\nabla)\ln\gamma+\alpha(\rho+p)\gamma^2(\textbf{V}.\nabla
\textbf{V}).\textbf{v}-\alpha(\rho+p)(\textbf{v}.\nabla \ln u)=0,\\
\label{4.6} &&\{(\rho+p)\gamma^2 \delta_{ij}+(\rho+p)\gamma^4 V_i
V_j\}\frac{1}{\alpha} \frac{\partial v^j}{\partial t}+(\rho+p)
\gamma^2v_{i,j}V^j +(\rho+p)\gamma^4 V_i v_{j,k} V^jV^k\nonumber\\
&&=-\gamma^2\{(\delta \rho+\delta p)+2(\rho+p)\gamma^2
(\textbf{V}.\textbf{v})\}a_i-(\delta p)_{,i}
-(\rho+p)\gamma^4(v_i V^j + v^j V_i)V_{k,j}V^k\nonumber\\
&&-\gamma^2\{(\delta \rho+\delta p)V^j
+2(\rho+p)\gamma^2(\textbf{V}.\textbf{v})V^j+(\rho+p)v^j\} V_{i,j}\nonumber\\
&&-\gamma^4V_i\{(\delta \rho+\delta p)V^j+4(\rho+p)\gamma^2
(\textbf{V}.\textbf{v})V^j+(\rho+p)v^j\}V_{j,k}V^k,\\
\label{4.7} &&\gamma^2\frac{1}{\alpha}\frac{\partial
(\delta\rho+\delta p)}{\partial
t}+\frac{2}{\alpha}(\rho+p)\gamma^4\textbf{V}.
\frac{\partial\textbf{v}}{\partial
t}-2(\rho+p)\gamma^4(\textbf{V}.\textbf{v})(\textbf{V}.\nabla)\ln
u+6(\rho+p)\gamma^6(\textbf{V}.\textbf{v})\{\textbf{V}.(\textbf{V}.\nabla)\textbf{V}\}
\nonumber\\
&&+(\rho+p)\gamma^4\textbf{V}.(\textbf{v}.\nabla)\textbf{V}+2(\rho+p)\gamma^4
\{\textbf{V}.(\textbf{V}.\nabla)\textbf{v}+\textbf{v}.(\textbf{V}.\nabla)\textbf{V}\}
+2(\delta\rho+\delta
p)\gamma^2\textbf{V}.\textbf{a}\nonumber\\
&&+2(\rho+p)\gamma^4(\textbf{V}.\textbf{v})
(\textbf{V}.\textbf{a})+(\rho+p)\gamma^2(\nabla.\textbf{v})
+(\rho+p)\gamma^2\textbf{v}.\textbf{a}
+\gamma^2(\textbf{V}.\nabla)(\delta\rho+\delta p)
-\frac{1}{\alpha}\frac{\partial(\delta p)}{\partial t}\nonumber\\
&&-(\rho+p)\gamma^2 (\textbf{v}.\nabla)\ln u+2(\delta \rho+\delta
p)\gamma^4
\textbf{V}.(\textbf{V}.\nabla)\textbf{V}+(\delta\rho+\delta
p)\gamma^2(\nabla.\textbf{V})\nonumber\\
&&+2(\rho+p)\gamma^4(\textbf{V}.\textbf{v})(\nabla.\textbf{V})=0.
\end{eqnarray}
The component form of Eqs.(\ref{4.5})-(\ref{4.7}) is
\begin{eqnarray}\label{4.8}
&&\rho\frac{\partial\tilde{\rho}}{\partial t}+p\frac{\partial
\tilde{p}}{\partial t}+(\rho+p)\gamma^2\left\{V\frac{\partial
v_x}{\partial t}+u\frac{\partial v_z}{\partial t}\right\}+\alpha
u\rho\tilde{\rho}_{,z}+\alpha
up\tilde{p}_{,z}+\alpha(\rho+p)\{\gamma^2Vuv_{x,z}+(1+\gamma^2 u^2)v_{z,z}\}\nonumber\\
&&-\frac{1}{\gamma}(\tilde{\rho}-\tilde{p})(\alpha u \gamma
p)_{,z}+\alpha(\rho+p)\gamma^2 u\{(1+2\gamma^2
V^2)V'+2\gamma^2uVu'\}v_x\nonumber\\
&&-\alpha(\rho+p)\left\{(1-2\gamma^2u^2)(1+\gamma^2 u^2)\frac{u'}{u}
-2\gamma^4u^2VV'\right\}v_z=0,\\
\label{4.9} &&(\rho+p)\gamma^2(1+\gamma^2V^2)\frac{1}{\alpha}
\frac{\partial v_x}{\partial t}+(\rho+p)\gamma^4 uV\frac{1}{\alpha}
\frac{\partial v_z}{\partial t}+(\rho+p)\gamma^2u(1+\gamma^2V^2)
v_{x,z}+(\rho+p)\gamma^4u^2Vv_{z,z}\nonumber\\
&&=-(\rho\tilde{\rho}+p\tilde{p})\gamma^2u\{(1+\gamma^2V^2)V'+\gamma^2uVu'\}
-(\rho+p)\gamma^4u\{(1+4\gamma^2V^2)uu'+4VV'(1+\gamma^2V^2)\}v_x\nonumber\\
&&-(\rho+p)\gamma^2[\{(1+2\gamma^2u^2)(1+2\gamma^2V^2)-\gamma^2V^2\}V'
+2\gamma^2(1+2\gamma^2u^2)uVu']v_z,\\
\label{4.10} &&(\rho+p)\gamma^2(1+\gamma^2
u^2)\left(\frac{1}{\alpha} \frac{\partial v_z}{\partial t}+u
v_{z,z}\right)+(\rho+p)\gamma^4uV\left(\frac{1}{\alpha}
\frac{\partial v_x}{\partial t}+u v_{x,z}\right)\nonumber\\
&&=-(\rho\tilde{\rho}+p\tilde{p})\gamma^2\{a_z+(1+\gamma^2u^2)uu'+\gamma^2
u^2 VV'\}
-p'\tilde{p}-p\tilde{p}_{,z}\nonumber\\
&&-(\rho+p)[\gamma^4\{u^2V'(1+4\gamma^2u^2)+2V\{a_z+(1+2\gamma^2u^2)uu'\}\}]v_x\nonumber\\
&&-(\rho+p)\gamma^2[u'(1+\gamma^2u^2)(1+4\gamma^2u^2)
+2u\gamma^2\{a_z+(1+2\gamma^2u^2)VV'\}]v_z,\\
\label{4.11}
&&\gamma^2\frac{\rho}{\alpha}\frac{\partial\tilde{\rho}}{\partial
t}+(\gamma^2-1)\frac{p}{\alpha}\frac{\partial\tilde{p}}{\partial
t}+\frac{2}{\alpha}(\rho+p)\gamma^4\left(V\frac{v_x}{\partial
t}+u\frac{v_z}{\partial
t}\right)+2(\rho+p)\gamma^4uVv_{x,z}+(\rho+p)\gamma^2(1+2\gamma^2u^2)v_{z,z}\nonumber\\
&&+\tilde{\rho}[2\rho\gamma^2u\{a_z+\gamma^2(VV'+uu')\}+\gamma^2u\rho']
+\tilde{p}[2p\gamma^2u\{a_z+\gamma^2(VV'+uu')\}+\gamma^2up']\nonumber\\
&&+\gamma^2u\rho\tilde{\rho}_{,z}+\gamma^2u'\rho\tilde{\rho}
+\gamma^2up\tilde{p}_{,z}+\gamma^2u'p\tilde{p}
+2(\rho+p)\gamma^4\{3\gamma^2uV(uu'+VV')+uVa_z+u'V\}v_x\nonumber\\
&&+(\rho+p)\gamma^2\left\{6\gamma^4u^2(VV'+uu')+\gamma^2(VV'+uu')
+a_z(1+2\gamma^2u^2)-\frac{u'}{u}+2\gamma^2uu'\right\}=0.
\end{eqnarray}

\renewcommand{\theequation}{C\arabic{equation}}
{\setcounter{equation}{0}}
\section*{Appendix C}

The GRMHD equations for the rotating magnetized background with
isothermal state of plasma remain the same as given by
Eqs.(\ref{3})-(\ref{7}). Hence their perturbed form will remain the
same as given in Appendix A (i.e., Eqs.(\ref{3.2})-(\ref{3.6})).

In this case, fluid's four-velocity is the same as given in the
previous section. The rotating magnetic field can be expressed in
$xz$-plane, i.e.,
$\textbf{B}=B[\lambda(z)\textbf{e}_\textbf{x}+\textbf{e}_\textbf{z}].$
We shall use the following notations for the perturbed magnetic
field in addition to the notations given by Eq.(\ref{4.4})
\begin{eqnarray*}
\textbf{b}\equiv \frac{\delta
\textbf{B}}{B}=b_x(t,z)\textbf{e}_\textbf{x}+b_z(t,z)\textbf{e}_\textbf{z}.
\end{eqnarray*}

Hence, the component form of Eqs.(\ref{3.2})-(\ref{3.6}) can be
written, after a tedious algebra, as follows
\begin{eqnarray}\label{5.7}
&&\frac{\partial b_x}{\partial t}+\alpha
ub_{x,z}=\alpha'(v_x-\lambda v_z +V b_z-u
b_x)+\alpha(v_{x,z}-\lambda v_{z,z}-\lambda' v_{z}+V'b_z+Vb_{z,z}-u'b_x),\\
\label{5.8}
&&\frac{\partial b_z}{\partial t}+\alpha ub_{z,z}=0,\\
\label{5.9}
&&b_{z,z}=0,\\
\label{5.10} &&\rho\frac{\partial\tilde{\rho}}{\partial
t}+p\frac{\partial \tilde{p}}{\partial
t}+(\rho+p)\gamma^2\left\{V\frac{\partial v_x}{\partial
t}+u\frac{\partial v_z}{\partial t}\right\}+\alpha
u\rho\tilde{\rho}_{,z}+\alpha
up\tilde{p}_{,z}+\alpha(\rho+p)\{\gamma^2Vuv_{x,z}+(1+\gamma^2
u^2)v_{z,z}\}\nonumber\\
&&-\frac{1}{\gamma}(\tilde{\rho}-\tilde{p})(\alpha u \gamma
p)_{,z}+\alpha(\rho+p)\gamma^2 u\{(1+2\gamma^2
V^2)V'+2\gamma^2uVu'\}v_x\nonumber\\
&&-\alpha(\rho+p)\left\{(1-2\gamma^2u^2)(1+\gamma^2 u^2)\frac{u'}{u}
-2\gamma^4u^2VV'\right\}v_z=0,\\
\label{5.11} &&\left\{(\rho+p)
\gamma^2(1+\gamma^2V^2)+\frac{B^2}{4\pi}\right\}\frac{1}{\alpha}
\frac{\partial v_x}{\partial t}+\left\{(\rho+p) \gamma^4 u
V-\frac{\lambda B^2}{4\pi}\right\}\frac{1}{\alpha}\frac{\partial
v_z}{\partial t}\nonumber
\end{eqnarray}
\begin{eqnarray}
&&+\left\{(\rho+p)\gamma^2(1+\gamma^2 V^2)-\frac{B^2}{4\pi}\right\}u
v_{x,z}+\left\{(\rho+p)\gamma^4uV+\frac{\lambda
B^2}{4\pi}\right\}uv_{z,z}-\frac{B^2}{4\pi}(1-u^2)b_{x,z}
\nonumber\\&&-\frac{B^2}{4\pi\alpha}b_x \{\alpha'(1-u^2)-\alpha
uu'\}+(\tilde{\rho}\rho+p\tilde{p})\gamma^2 u\{(1+\gamma^2
V^2)V'+\gamma^2uVu'\}\nonumber\\
&&+\left[(\rho+p)\gamma^4u\{(1+4\gamma^2 V^2)u
u'+4(1+\gamma^2V^2)VV'\}+\frac{B^2 u \alpha'}{4\pi \alpha}\right]v_x\nonumber\\
&&+\left[(\rho+p)\gamma^2 [\{(1+2 \gamma^2 u^2)(1+2 \gamma^2 V^2) -
\gamma^2 V^2\}V'+2\gamma^2(1+2\gamma^2 u^2)u V u'] +\frac{B^2 u}{4
\pi\alpha}(\lambda\alpha)'\right]v_z=0,\\
\label{5.12}
&&\left\{(\rho+p) \gamma^2(1+\gamma^2 u^2)+ \frac{\lambda^2
B^2}{4\pi}\right\}\frac{1}{\alpha}\frac{\partial v_z}{\partial
t}+\left\{(\rho+p) \gamma^4 u V-\frac{\lambda
B^2}{4\pi}\right\}\frac{1}{\alpha}\frac{\partial v_x}{\partial t}
\nonumber\\
&&+\left\{(\rho+p) \gamma^2(1+\gamma^2 u^2) -\frac{\lambda^2
B^2}{4\pi}\right\}u v_{z,z}+\left\{(\rho+p)\gamma^4 u
V+\frac{\lambda B^2}{4\pi})uv_{x,z}\right\}+\frac{\lambda B^2}{4
\pi}(1-u^2)b_{x,z}\nonumber\\
&&+\frac{B^2}{4\pi \alpha}\{-(\alpha
\lambda)'+\alpha'\lambda-u\lambda(u\alpha)'\}b_x+(\rho\tilde{\rho}+p\tilde{p})
\gamma^2[a_z+u\{(1+\gamma^2 u^2)u'+\gamma^2uVV'\}]\nonumber\\
&&+\left[(\rho+p)\gamma^4\{u^2V'(1+4\gamma^2V^2)+2V(a_z+uu'(1+2\gamma^2
u^2))\}+\frac{\lambda B^2 \alpha'u}{4\pi\alpha}\right]v_x\nonumber\\
&&+\left[(\rho+p)\gamma^2\{u'(1+\gamma^2 u^2)(1+4\gamma^2 u^2)
+2u\gamma^2\{(1+2\gamma^2 u^2)VV'+a_z\}\}-\frac{\lambda B^2
u}{4\pi\alpha}(\alpha\lambda)'\right]v_z=0,\\
\label{5.13} &&\gamma^2\rho\frac{1}{\alpha}\frac{\partial
\tilde{\rho}}{\partial
t}+(\gamma^2-1)p\frac{1}{\alpha}\frac{\partial \tilde{p}}{\partial
t}+\frac{B^2}{4\pi\alpha}(u\lambda-V)(u\frac{\partial b_x}{\partial
t}-V\frac{\partial b_z}{\partial t}+\lambda\frac{\partial
v_z}{\partial t}-\frac{\partial
v_x}{\partial t})\nonumber\\
&&+\frac{2}{\alpha}(\rho+p)\gamma^4(V\frac{\partial v_x}{\partial
t}+u\frac{\partial v_z}{\partial
t})+2(\rho+p)\gamma^4Vuv_{x,z}+(\rho+p)\gamma^2(1+2\gamma^2u^2)v_{z,z}\nonumber\\
&&+\tilde{\rho}\gamma^2[2\rho u\{a_z+\gamma^2(VV'+uu')\}+\rho'u+\rho
u']+\tilde{p}\gamma^2[2pu\{a_z+\gamma^2(VV'+uu')\}+p'u+p
u']\nonumber\\
&&+\gamma^2u\rho\tilde{\rho}_{,z}+\gamma^2up\tilde{p}_{,z}
-\frac{B^2V}{4\pi\alpha}(\alpha\lambda)'b_z+\frac{B^2u}{4\pi\alpha}(\alpha\lambda)'b_x
+\frac{B^2(u\lambda-V)}{4\pi\alpha}(\alpha
b_{x,z}+\alpha'b_x)\nonumber\\
&&+v_z\left[(\rho+p)\gamma^2\left\{6\gamma^4u^2(VV'+uu')+\gamma^2(VV'+uu')
+(1+2\gamma^2u^2)a_z-\frac{u'}{u}+2\gamma^2uu'\right\}
+\frac{B^2\lambda}{4\pi\alpha}(\alpha\lambda)'\right]\nonumber\\
&&+v_x\left[(\rho+p)\gamma^4\{2uVa_z+6\gamma^2uV(VV'+uu')+2uV'\}
-\frac{B^2}{4\pi\alpha}(\alpha\lambda)'\right]=0.
\end{eqnarray}

\end{document}